\documentclass[11pt,a4paper, final, twoside]{article}
\usepackage[centertags]{amsmath}
\usepackage{amsfonts}
\usepackage{amssymb}
\usepackage{amsthm}
\usepackage{newlfont}
\usepackage{fancyhdr}
\usepackage{amscd}
\usepackage{sectsty}
\usepackage{siunitx}
\usepackage{graphicx}
\usepackage{subfigure}
\usepackage{cite}
\usepackage{etoolbox}
\usepackage{caption}
\usepackage{array}
\usepackage{booktabs}
\usepackage[table]{xcolor}
\usepackage{xcolor,colortbl}
\usepackage{colortbl,multirow,hhline}
\usepackage{afterpage}
\definecolor{blue(ncs)}{rgb}{0.0, 0.53, 0.74}
\definecolor{blizzardblue}{rgb}{0.80, 0.91, 0.90}
\usepackage[colorlinks=true, urlcolor=blue(ncs),  linkcolor=blue(ncs), citecolor=blue(ncs)]{hyperref}
\usepackage{latexsym,epstopdf,float}
\usepackage{euscript,pifont,mathrsfs,latexsym,graphicx}

\newcommand{\midsepremove}{\aboverulesep = 0.31mm \belowrulesep = 0.31mm}\midsepremove
\newcommand{\midsepdefault}{\aboverulesep = 0.605mm \belowrulesep = 0.984mm}\midsepdefault

\sectionfont{\color{blue(ncs)}}
\subsectionfont{\color{blue(ncs)}}
\patchcmd\thebibliography{\labelsep}{\labelsep\itemsep=0pt\parsep=0pt\relax}{}{\typeout{Couldn't patch the command}}
\renewenvironment{abstract}
{
\begin{center}%
\end{center}%
{\color{blue(ncs)}\normalfont\textbf{ABSTRCT:}}
}

\numberwithin{equation}{section}

\begin{document}

\title{
The effects of beta-cell mass and function, intercellular coupling, and islet synchrony on  $\textrm{Ca}^{2+}$  dynamics
}

\author{
Maryam Saadati$^1$, Yousef Jamali$^{^1*}$
}
\date{%
    $^1$ Biomathematics Laboratory, Department of Applied Mathematics, School of Mathematical Sciences, Tarbiat Modares University,Tehran, Iran\\
    $^*$ Corresponding author. Email: y.jamali@modares.ac.ir \\[2ex]%
    \today
}

\maketitle
\begin{abstract}
Type 2 diabetes (T2D) is a challenging metabolic disorder characterized by a substantial loss of $\beta$-cell mass and alteration of $\beta$-cell function in the islets of Langerhans, disrupting insulin secretion and glucose homeostasis. The mechanisms for deficiency in $\beta$-cell mass and function during the hyperglycemia development and T2D pathogenesis are complex. To study the relative contribution of $\beta$-cell mass to $\beta$-cell function in T2D, we make use of a comprehensive electrophysiological model of human $\beta$-cell clusters. We find that defect in $\beta$-cell mass causes a functional decline in single $\beta$-cell, impairment in intra-islet synchrony, and changes in the form of oscillatory patterns of membrane potential and intracellular $\textrm{Ca}^{2+}$ concentration, which can lead to changes in insulin secretion dynamics and in insulin levels. The model demonstrates a good correspondence between suppression of synchronizing electrical activity and published experimental measurements. We then compare the role of gap junction-mediated electrical coupling with both $\beta$-cell synchronization and metabolic coupling in the behavior of $\textrm{Ca}^{2+}$ concentration dynamics within human islets. Our results indicate that inter-$\beta$-cellular electrical coupling depicts a more important factor in shaping the physiological regulation of islet function and in human T2D. We further predict that varying the whole-cell conductance of delayed rectifier $\textrm{K}^{+}$ channels modifies oscillatory activity patterns of $\beta$-cell population lacking intercellular coupling, which significantly affect $\textrm{Ca}^{2+}$ concentration and insulin secretion.
\end{abstract}

\afterpage{} \fancyhead{} \fancyfoot{} \fancyhead[LE, RO]{\bf\thepage}

\section*{Introduction}\hypertarget{Intro}{}
The human pancreatic $\beta$-cells placed in islets of Langerhans organize a complex functional network \cite{3,1,2,4}, and ensure blood glucose homeostasis \cite{91} through a pulsatile, well-regulated insulin secretion \cite{87,5,6}. Like their rodent counterparts \cite{83,7,8}, human $\beta$-cells respond to increasing plasma glucose concentrations with acceleration of metabolism and elevation in ATP levels, which in turn inhibits ATP-sensitive potassium ($\textrm{K}_{\textrm{ATP}}$) channels, leading to membrane depolarization, activation of voltage-dependent calcium channels (VDCCs), a rise in the cytosolic calcium concentration ($[\textrm{Ca}^{2+}]_{\textrm{c}}$), and triggering exocytosis of insulin granules \cite{84,11,9,10}.

Both murine and human dissociated $\beta$-cells are intrinsically and functionally heterogeneous with respect to biophysical characteristics, metabolic responses, electrical dynamics, and insulin release \cite{85,86,12,13}. Within rodent islets, gap junctions consisting of the Connexin36 (Cx36) protein form a strong intercellular electrical coupling between heterogeneous $\beta$-cells, which is potentially important for coordination of oscillations in the $\beta$-cell intracellular $\textrm{Ca}^{2+}$ and insulin secretion across the islet, enhancing pulsatility of insulin secretion and regulating glucose homeostasis \cite{15,16,17,104,18}. Similarly, in human pancreatic islets Cx36 gap junctions help to overcome the heterogeneity of individual $\beta$-cells, and give rise to bursting behavior in synchrony upon glucose stimulation \cite{18,19,20}. Additionally, Cx36 gap junction electrical coupling mediates a marked suppression of spontaneous $[\textrm{Ca}^{2+}]_{\textrm{c}}$ elevations at basal glucose, generates a characteristic sigmoidal secretory response to increasing glucose, enhances the peak amplitude of first-phase insulin release, and coordinates the pulsatile second-phase insulin release \cite{88,34,22}, which in turn are important for glucose homeostasis. These observations illustrate a strong link between glucose-stimulated insulin secretion and gap junction function. Recent investigations have indicated the presence of heterogeneity in gap junctional conductance between the heterogeneous $\beta$-cells \cite{17,23,105} that influences the complex functional organization of $\beta$-cells \cite{82,24,25} and the spatiotemporal characteristics of $\textrm{Ca}^{2+}$ waves under the islet mathematical models \cite{92,29,26,106}.

Besides electrical communications, metabolic communications are mediated by gap junction channels. These intercellular channels permit cell-to-cell diffusion of specific signaling ions and some glycolytic intermediates, which strongly affects the pattern of $[\textrm{Ca}^{2+}]_{\textrm{c}}$ oscillations and insulin secretory profiles \cite{28}. Intact islets display electrical behaviour consisting of so-called slow bursting pattern with a period of $\sim 5$ min, corresponding to the frequency of metabolism, $[\textrm{Ca}^{2+}]_{\textrm{c}}$, and insulin oscillations \cite{31,32}. It is believed that this rhythmic islet activity is due to slow oscillatory dynamics of underlying glycolysis and metabolism observed in mouse and human $\beta$-cells \cite{95,33}. In particular, glycolytic oscillations are proposed to be the key player in the overall islet activity at stimulatory glucose concentrations.

Cx36 knockout mice display a reduction in the synchrony of $[\textrm{Ca}^{2+}]_{\textrm{c}}$ oscillations and disruption of calcium wave propagation resulting in impaired pulsatile patterns of insulin release and glucose intolerance \cite{34,26,27}. Importantly, these changes in $\textrm{Ca}^{2+}$ activity and insulin secretion dynamics have also been observed in the prediabetes stage and patients with type 2 diabetes \cite{93,35,36,37,38}. This, along with mouse models of prediabetes which have demonstrated Cx36 disruption characteristic at this stage \cite{39}, suggest that changes in Cx36-mediated coupling may be a key determining factor in islet dysfunction and development of T2D \cite{16,40}. Previous reports have shown that the expression of Cx36 protein is decreased in islets exposed to chronic hyperglycemia \cite{14}. Altered Cx36 gap junction function makes the pancreatic islets more sensitive to $\beta$-cell damage and lower efficiency in insulin secretion \cite{34,41}, which again suggests a potential role for decreased coupling \mbox{in T2D characterized by progressive $\beta$-cell death.}

Plasma insulin levels depend on the absolute number of insulin secreting $\beta$-cells (i.e., $\beta$-cell mass) and the functional status of each of these cells (i.e., $\beta$-cell function). Hereby, deficiency in either dynamics of $\beta$-cell mass or function, or both, results in insulin insufficiency and the onset of hyperglycemia. Current research in diabetes reveals that besides a substantial decline in $\beta$-cell mass, a significant defect in $\beta$-cell function is evident in T2D patients \cite{94,44,43,42}. Furthermore, other studies indicate that at the time of T2D diagnosis diabetic islets seem to have lost $\sim 50 \%$ of their $\beta$-cells \cite{43,42,45}, which is tightly correlated with islet dysfunction including a reduction in the amplitude of first-phase insulin secretion and impairment in the secretory pulses during second-phase insulin secretion \cite{33,37}. Clinically, the most common therapeutic approaches for T2D aim to regenerate $\beta$-cell mass or to preserve $\beta$-cell function. Addressing the latter needs a deep understanding of the contribution and kinetics of $\beta$-cell mass and function in T2D etiology and pathogenesis. Thus, in the present study, we examined whether $\beta$-cell defects are intrinsically functional or whether a reduction in $\beta$-cell mass is linked to $\beta$-cell dysfunction. We first constructed a multicellular computational model of heterogeneous and heterogeneously coupled $\beta$-cells, and analyzed the effects on functional behavior of single $\beta$-cell caused by changes in $\beta$-cell population of human islets. We obtained a nice agreement between theoretical results and experimental data regarding the disruption in normal oscillatory patterns of insulin secretion after $\sim 50 \%$ $\beta$-cell loss. We then compared the behavior of electrical activity and $[\textrm{Ca}^{2+}]_{\textrm{c}}$ dynamics after reduction in gap junction coupling. By combining the effect of varying coupling strengths and glucose stimulations, we investigated how $[\textrm{Ca}^{2+}]_{\textrm{c}}$ levels altered with loss of $0 \%$ and $\sim 55 \%$ $\beta$-cell mass. Finally, we predicted the impact of delayed rectifier $\textrm{K}^{+}$ (Kv) channels on the electrical behavior of uncoupled $\beta$-cell population.

\section*{Computational methods}\hypertarget{CompMeth}{}

\subsection*{Model of $\beta$-cell}

The Hodgkin-Huxley type model for human $\beta$-cells has been developed by Pedersen \cite{49}, who carefully described the electrophysiological properties of ion channels in human $\beta$-cells, and then Riz et al. included $\textrm{Ca}^{2+}$ dynamics in the model \cite{50}. For this study, we prefer to use such a formulation because it provides a firm explanation for the human $\beta$-cell dynamics, confirmed by experimental investigations \cite{51,52}. The model is composed of an electrical component and a glycolytic component \cite{103}. It includes membrane potential activity, cytosol and submembrane dynamics of $\textrm{Ca}^{2+}$, and glucose metabolism. The glycolytic oscillatory component drives slow bursting patterns \cite{49,50,53}, underlying slow $\textrm{Ca}^{2+}$ oscillations and insulin release pulses \cite{31,32}.

Briefly, the membrane potential $(V_{\textrm{i}})$ of a single $\beta$-cell i follows:
\begin{equation*}
\dfrac{dV_{\textrm{i}}}{dt}=-(I_{\textrm{SK}} + I_{\textrm{BK}} + I_{\textrm{Kv}} + I_{\textrm{HERGA}} + I_{\textrm{Na}} + I_{\textrm{CaL}} + I_{\textrm{CaPQ}} + I_{\textrm{CaT}} + I_{\textrm{K(ATP)}} + I_{\textrm{leak}}),
\end{equation*}
where $I_{\textrm{X}}$ denotes the transmembrane current conducted by channel type $\textrm{X}$. Full equations and parameters of the model can be found in the supplementary material.

\subsection*{Network of $\beta$-cells}

There exists evidence that the mean $\beta$-cell number for each human pancreatic islet is $\sim 10^3$ \cite{54,55,56}. For lattice structure of islet, we model cubic network including $10 \times 10 \times 10$ $\beta$-cells such that each central cell is surrounded by 6 neighbors (the $3-\textrm{D}$ Von Neumann neighborhoods of cellular automata theory). These cells are coupled with adjacent $\beta$-cells through both electrical and metabolic connections. The equation for membrane potential of the ith $\beta$-cell surrounded by j neighboring cells is modified to simulate electrical gap junction coupling in the modeled islet:
\begin{equation*}
  \dfrac{dV_{\textrm{i}}}{dt}=- I_{\textrm{ion,i}}- \sum_{\textrm{j}\in\Omega(\textrm{i})} g_{\textrm{c}}^{(\textrm{i,j})} (V_{\textrm{i}}-V_{\textrm{j}}),
\end{equation*}
where $g_{\textrm{c}}^{(\textrm{i,j})}$ refers to the electrical coupling conductance between cells i and j, and $\Omega(\textrm{i})$ is all adjacent cells of cell i.

As in another modeling study \cite{57}, to account for the metabolic coupling among $\beta$-cells we consider the diffusion of glucose-6-phosphate (G6P) between cells, which is assumed to be in rapid equilibrium with fructose-6-phosphate (F6P). The equation for the total concentrations of G6P and F6P in the ith $\beta$-cell surrounded by j neighboring cells is defined by:
\begin{equation*}
\dfrac{d\,G6P.F6P_{\textrm{i}}}{dt}=V_{\textrm{GK,i}}-V_{\textrm{PFK}}-P_{\textrm{G6P.F6P}}\sum_{\textrm{j}\in\Omega(\textrm{i})}(G6P.F6P_{\textrm{i}}-G6P.F6P_{\textrm{j}}),
\end{equation*}
where $V_{\textrm{GK,i}}$ is the glucokinase reaction rate of cell i, $V_{\textrm{PFK}}$ refers to the phospho-fructokinase reaction rate, and the parameter $P_{\textrm{G6P.F6P}}$ describes the metabolic coupling strength computed by analyzing data of diffusion of glycolytic metabolites among the islet $\beta$-cells.

\subsection*{Numerical methods}

All equations of the model are written and implemented in a Python algorithm, and the forth-order Runge-Kutta numerical scheme is used for solving the ODE systems, both electrical and metabolic components with a time-step of 0.02 and 0.05 ms, respectively.

In the present study, biological heterogeneity of the human $\beta$-cells is introduced by some crucial parameters that control the electrical behavior of the modeled $\beta$-cells. Specifically, cellular heterogeneity is represented in the conductances of the gap junction channels $(g_\textrm{c})$ and the delayed rectifier $\textrm{K}^{+}$ channels $(g_\textrm{Kv})$, and the maximal reaction rate of glucokinase enzyme $(V_\textrm{GK,max})$, which has strong effects on the slow oscillation frequency, and is also connected to heterogeneity glucose sensitivity. The values of parameters $g_\textrm{c}$ (and similarly $g_\textrm{Kv}$) and $V_\textrm{GK,max}$ select from normal distributions with mean value equal to original value of the parameters and standard deviation is set to $4 \%$ and $25 \%$ of the mean, respectively.

\section*{Results}\hypertarget{Res}{}

\subsection*{1. Does $\beta$-cell loss of mass primarily cause functional $\beta$-cell defect in type 2 diabetes?}\hypertarget{Res1}{}

\begin{figure}[!ht]
\centering
\subfigure
{
\includegraphics[width=12cm]{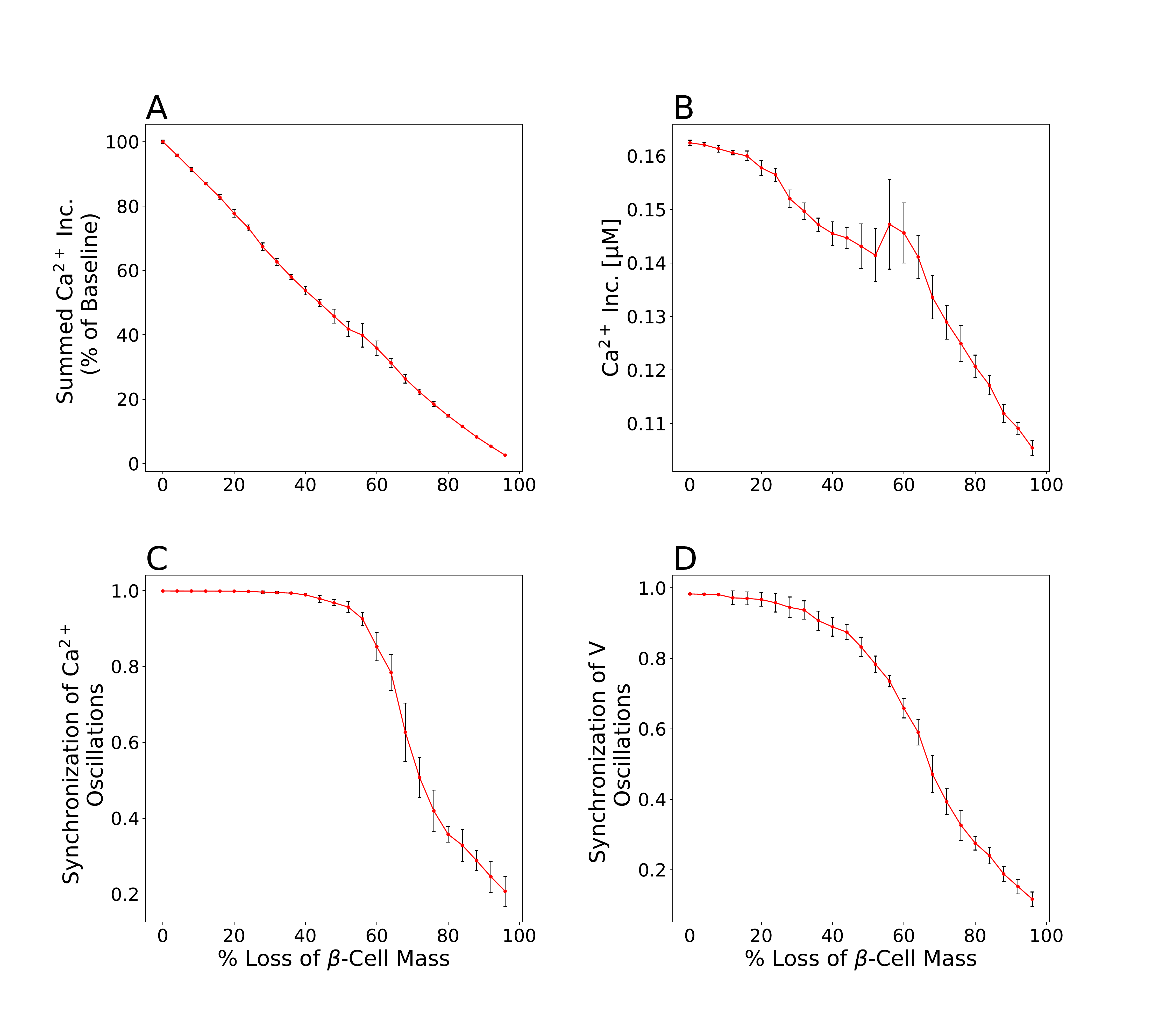}
}
\\
\vspace{-10mm}
\subfigure{
\includegraphics[width=12cm]{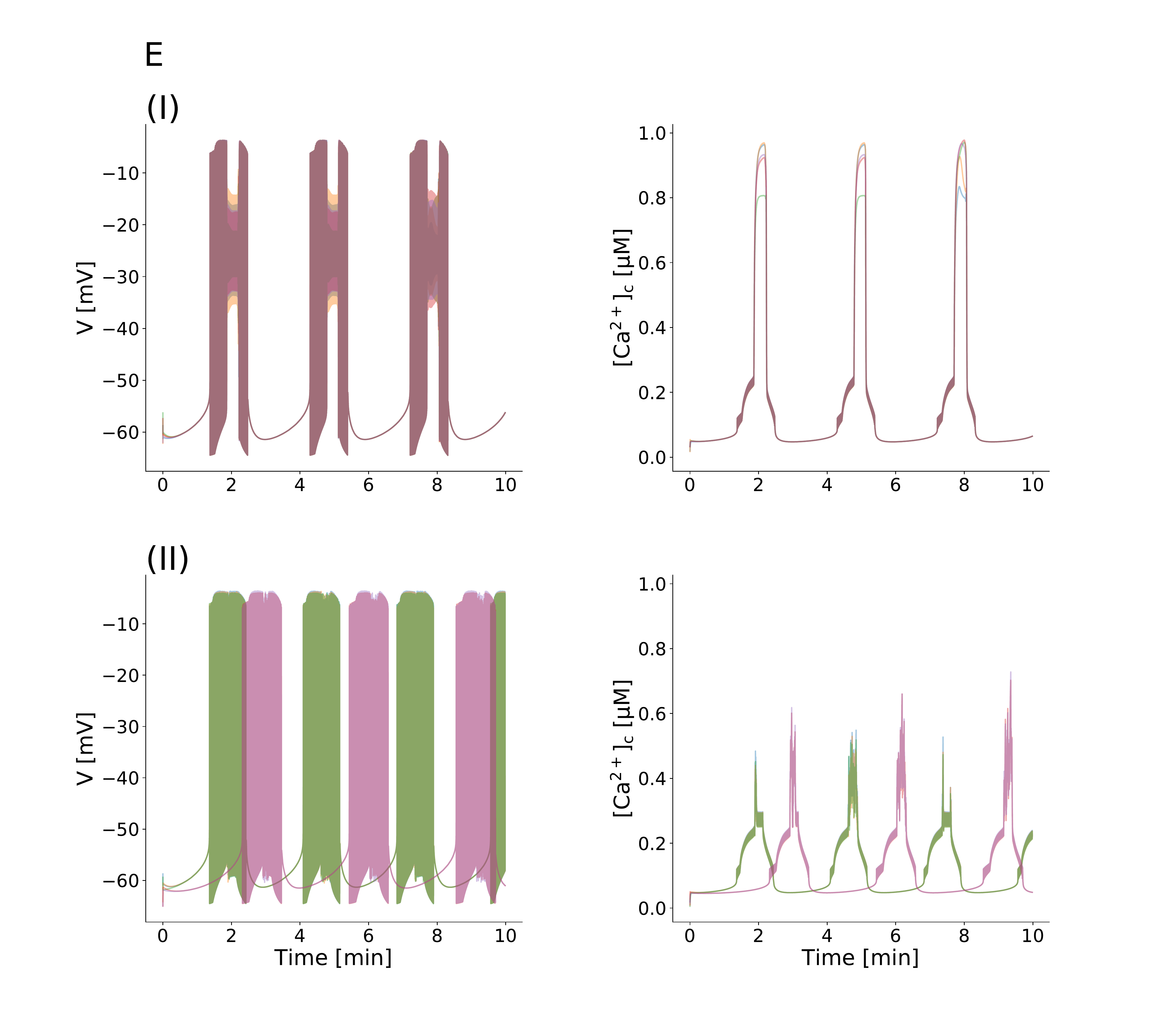}
}
\vspace{-6mm}
\caption{\scriptsize {\bf Loss of $\beta$-cell mass causes defect of $\beta$-cell function. (A)} Linear decrease in the summed $[\textrm{Ca}^{2+}]_{\textrm{c}}$ activity during reduction of $\beta$-cell mass. The number of cells reduced is represented as the $\%$ of all cells in the modeled islet (1000 cells). {\bf(B)} Mean $[\textrm{Ca}^{2+}]_{\textrm{c}}$ activity during reduction of $\beta$-cell mass. Note that deficiency in $\beta$-cell mass has the strong effect on $[\textrm{Ca}^{2+}]_{\textrm{c}}$ level. {\bf(C)},{\bf(D)} Dependence of the synchronization of $[\textrm{Ca}^{2+}]_{\textrm{c}}$ and membrane potential oscillations as a function of the number of cells reduced ($\%$ of islet). The intra-islet synchronization is disrupted after loss of $\sim 50\%$ $\beta$-cell mass. {\bf(E)} Representative simulated membrane potential and $[\textrm{Ca}^{2+}]_{\textrm{c}}$ time-courses across the islet lacking $10\%$ $\beta$-cells (subpanel {\bf(I)}) and lacking $80\%$ $\beta$-cells (subpanel {\bf(II)}). A massive decline in $\beta$-cell mass causes changes in the shape of electrical bursting. Data in {\bf(A)}-{\bf(D)} represent the mean $\pm$ S.E.M. over 10 modeled islets.}
\label{fig1}
\end{figure}

To investigate the contribution of $\beta$-cell mass and function to the insufficient insulin release and progression of T2D, we eliminated the modeled $\beta$-cells in islet network randomly, in order to more closely mimic the clinical conditions. We observed that with removing the simulated cells, not only the summed $[\textrm{Ca}^{2+}]_{\textrm{c}}$ activity was decreased (Fig.\,\ref{fig1}\,\textcolor{blue(ncs)}{A}) but surprisingly, $[\textrm{Ca}^{2+}]_{\textrm{c}}$ of single $\beta$-cell was reduced $\sim 32\%$ (Fig.\,\ref{fig1}\,\textcolor{blue(ncs)}{B}). These changes in intracellular $\textrm{Ca}^{2+}$ levels of an active $\beta$-cell were caused by substantial changes in the shape of electrical activity, in addition to impairment of coordinated electrical behavior in $\beta$-cell islet. When $\sim 10\%$ of the islet cells were lost, the pattern of oscillatory membrane potential, which correlated with the pattern of $[\textrm{Ca}^{2+}]_{\textrm{c}}$ oscillations, was noteworthy different from the deletion of $\sim 80\%$ $\beta$-cell mass (Fig.\,\ref{fig1}\,\textcolor{blue(ncs)}{E} (I and II)). In fact, the peak level of $[\textrm{Ca}^{2+}]_{\textrm{c}}$ was significantly lower in $\sim 80\%$ than $\sim 10\%$ loss due to widely varying patterns of the membrane potential oscillations, which occurred at a major decline in $\beta$-cell mass. Therefore, changes in $\beta$-cell mass caused the functional alterations in single $\beta$-cell, leading to changes in insulin concentration and secretion dynamics. This result can support to the hypothesis that deficit in $\beta$-cell mass induces various abnormalities in single $\beta$-cell function observed in patients with type 2 diabetes.

Additionally, we surprisingly noticed that $\beta$-cell electrical activity across the islet, specially oscillations of $\beta$-cell $[\textrm{Ca}^{2+}]_{\textrm{c}}$, completely synchronized before loss of $\sim 50\%$ the cells, and then the intra-islet synchronization dropped with loss of $> 50 \%$ (Fig.\,\ref{fig1}\,\textcolor{blue(ncs)}{C} and \textcolor{blue(ncs)}{D}), resembling experimental recordings which disruption in pulsatile insulin secretion linked to impairment in coordination of $[\textrm{Ca}^{2+}]_{\textrm{c}}$ oscillations can be found after reduction of $\sim 50\%$ the $\beta$-cell mass in the pathogenesis of type 2 diabetes \cite{43,42,45}. On the other hand, the summed $[\textrm{Ca}^{2+}]_{\textrm{c}}$ activity was partly linearly decreased with deletion of $\beta$-cells (Fig.\,\ref{fig1}\,\textcolor{blue(ncs)}{A}). Therefore, considering that the phase transition occurred in the intra-islet synchrony and that this behavior was not observed in the summed $[\textrm{Ca}^{2+}]_{\textrm{c}}$ response, it seems that the lack of insulin pulsatility patterns is a more important factor in type 2 diabetes. These simulations confirm previous studies, which show that $\sim 50 \%$ loss of $\beta$-cell mass is a critical point \cite{43,45}.

\subsection*{2. Intercellular coupling is more effective than islet synchronization on $\textrm{Ca}^{2+}$ concentrations}\hypertarget{Res2}{}

\begin{figure}[!ht]
\centering
{
\includegraphics[width=12.3cm]{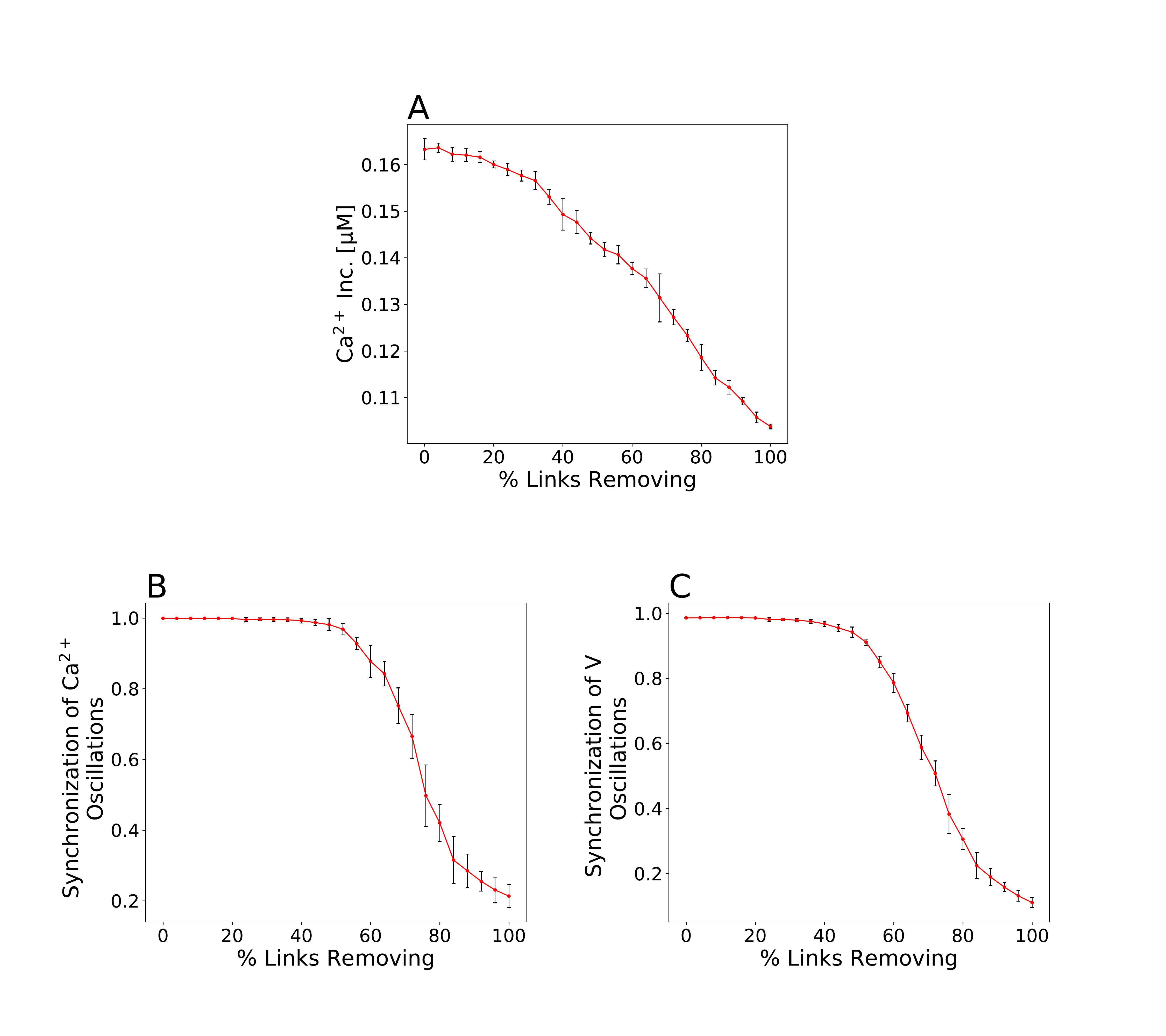}
}
\vspace{-3mm}
\caption{\scriptsize {\bf Gap junction coupling is more effective on $\textrm{Ca}^{2+}$ activity. (A)} Mean $[\textrm{Ca}^{2+}]_{\textrm{c}}$ response across the modeled islet as a function of fraction of gap junctions removed. Note that level of $[\textrm{Ca}^{2+}]_{\textrm{c}}$ decreases faster after loss of $\sim 50\%$ connections. {\bf(B)},{\bf(C)} The synchrony of $[\textrm{Ca}^{2+}]_{\textrm{c}}$ and membrane potential oscillations depends on fraction of gap junctions removed. Data represent the mean $\pm$ S.E.M. over 10 modeled islets.}
\label{fig2}
\end{figure}

We simulated slow oscillations with a period of $\sim 5$ min, and compared the influence of gap junction and synchronization on $[\textrm{Ca}^{2+}]_{\textrm{c}}$ levels of single human $\beta$-cell. To examine whether intercellular coupling or islet synchronization is more effective on the calcium dynamics, we remove randomly gap junction coupling between $\beta$-cells in the islet lattice (Fig.\,\ref{fig2}). With reduction in $\beta$-cell coupling, the average cytosolic calcium concentration began to fall, initially (Fig.\,\ref{fig2}\,\textcolor{blue(ncs)}{A}) while after deletion of $\sim 50\%$ gap junction connections, the synchrony of electrical dynamics across the islet started to reduce (Fig.\,\ref{fig2}\,\textcolor{blue(ncs)}{B} and \textcolor{blue(ncs)}{C}). In other words, as less than $\sim 50\%$ connections among islet $\beta$-cells were lost, $[\textrm{Ca}^{2+}]_{\textrm{c}}$ of single $\beta$-cell was gradually decreased whereas the whole-islet $\textrm{Ca}^{2+}$ activity remained synchronous. This result suggests that gap junctional conductance of  $\beta$-cell is a more important factor than islet synchrony of electrical patterns for intracellular $\textrm{Ca}^{2+}$ dynamics and levels of insulin secretion.

Moreover, Fig.\,\ref{fig2}\,\textcolor{blue(ncs)}{A} displayed the dual behavior of the islet $\beta$-cells: before deletion of $\sim 50\%$ gap junctions, the slope of average $[\textrm{Ca}^{2+}]_{\textrm{c}}$ was less than after deletion of $\sim 50\%$ intercellular connections. In fact, $[\textrm{Ca}^{2+}]_{\textrm{c}}$ level of single $\beta$-cell decreased substantially faster after a greater ($> 50 \%$) loss of cells compared with the loss of $< 50 \%$. These results indicate that a massive decline in the islet gap junction coupling, resulting in disruption of synchrony in human $\beta$-cell islet, plays a key role in more fast decrease in the average $[\textrm{Ca}^{2+}]_{\textrm{c}}$, poorly coordinated calcium dynamics, and plausibly, impairment of pulsatile insulin release and development of diabetes.

\subsection*{3. The interplay of gap junction electrical coupling and metabolic coupling in $[\textrm{Ca}^{2+}]_{\textrm{c}}$ activity}\hypertarget{Res3}{}

Loppini et al. considered metabolic coupling, in addition to electrical coupling, resulting in diffusion of glycolytic metabolites such as G6P and F6P molecules among the islet $\beta$-cells \cite{57}. We analyzed the role of electrical and metabolic coupling strength on the $\beta$-cell calcium activity by graded changes in the gap junction conductance and the metabolic coupling strength.

\begin{figure}[!ht]
\begin{tabular}{cc}

\hspace{-1cm}
\begin{minipage}{.5\textwidth}
\includegraphics[width=.79\textwidth]{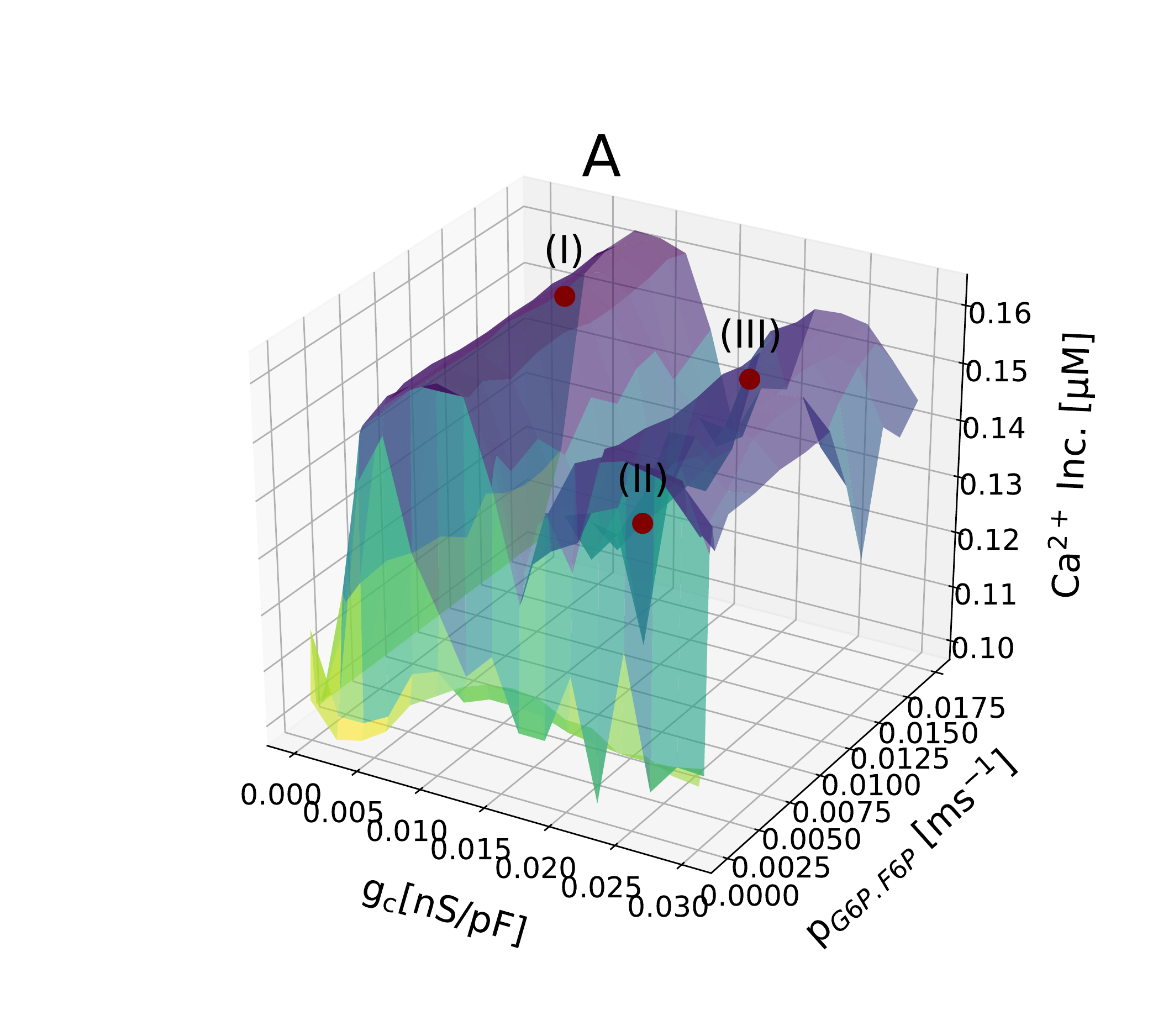}
\\
\includegraphics[width=.79\textwidth]{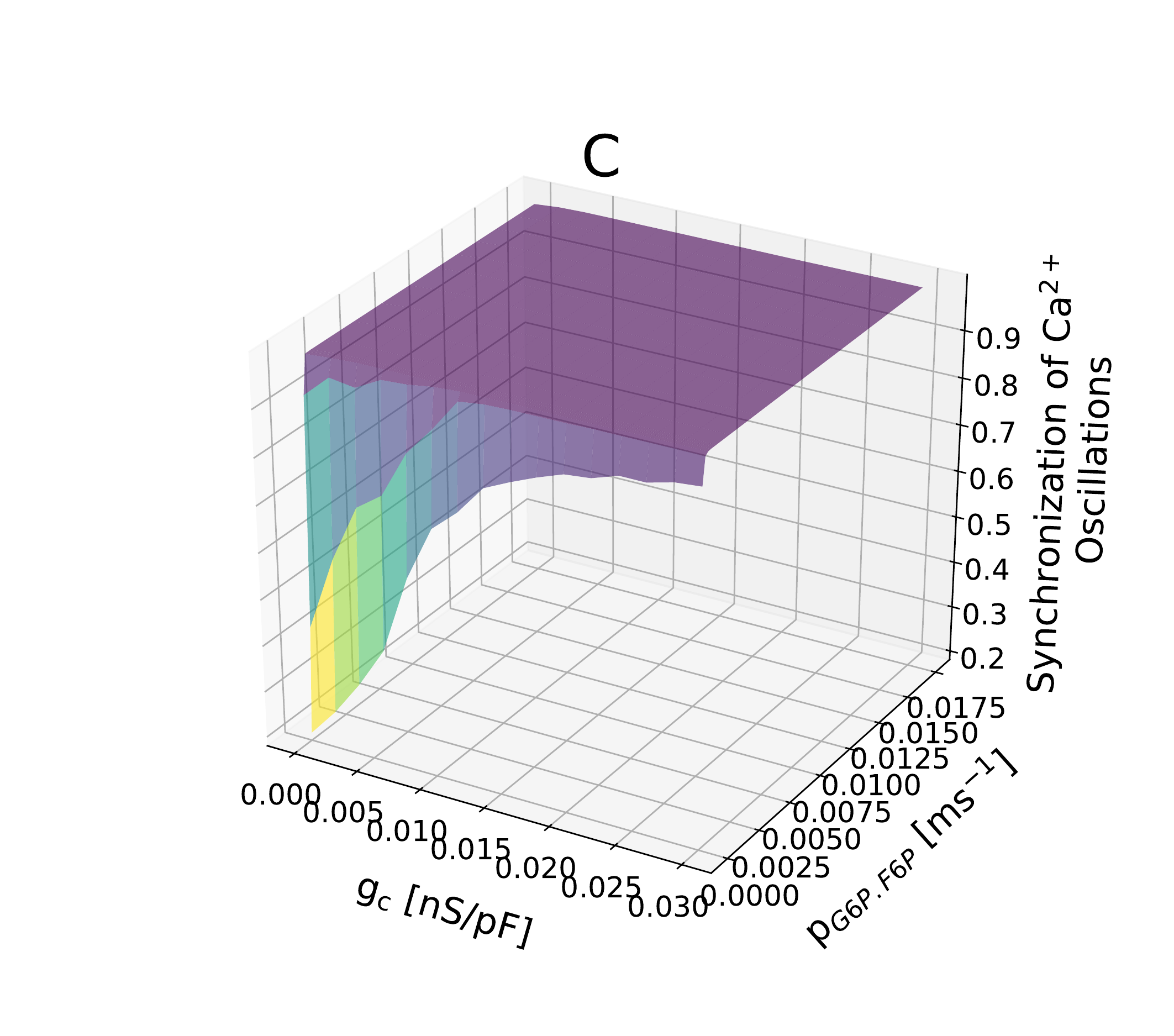}
\\
\includegraphics[width=.79\textwidth]{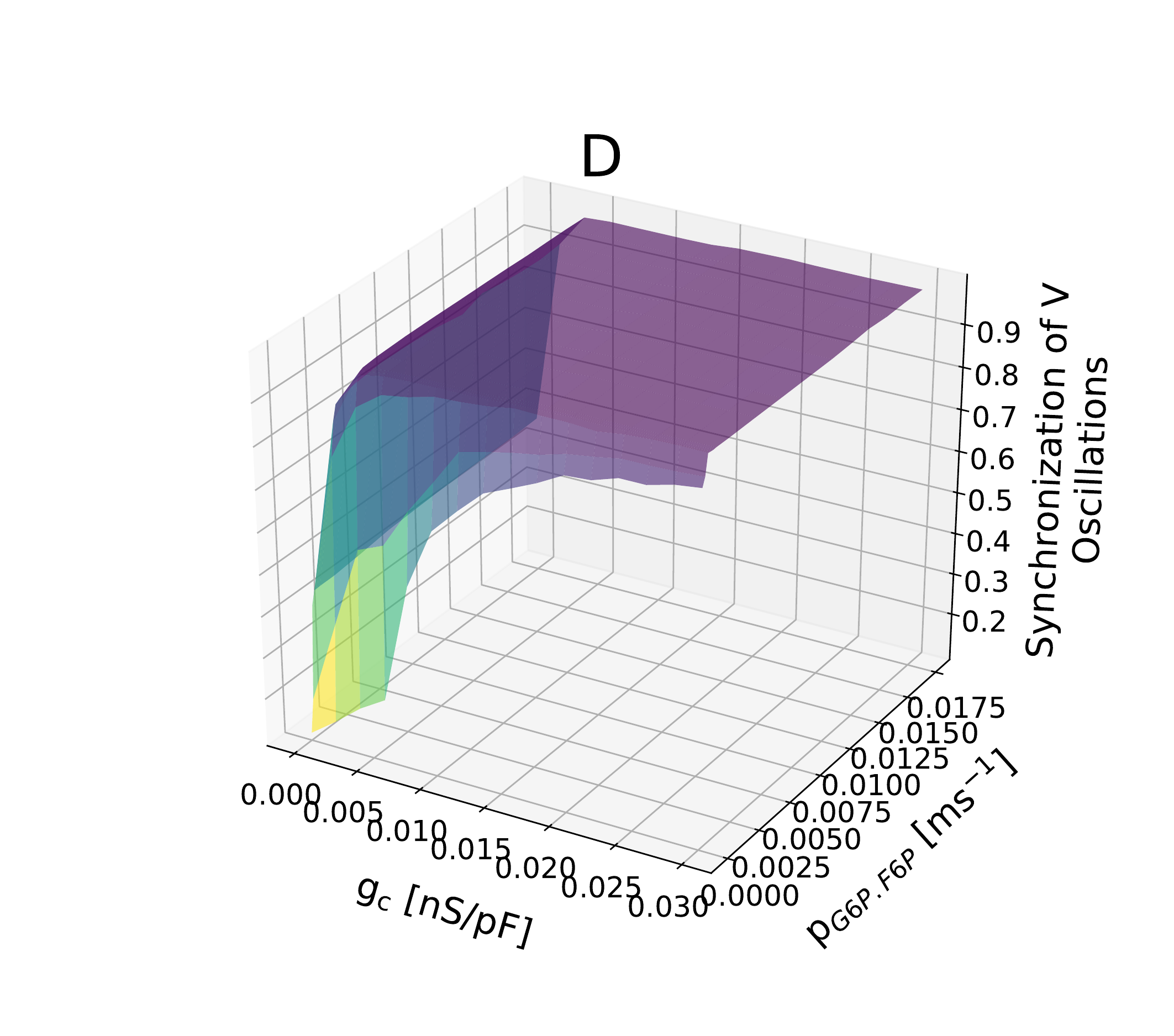}
\end{minipage}
&
\hspace{-1.9cm}
\begin{minipage}{.5\textwidth}
\includegraphics[width=1.3\textwidth]{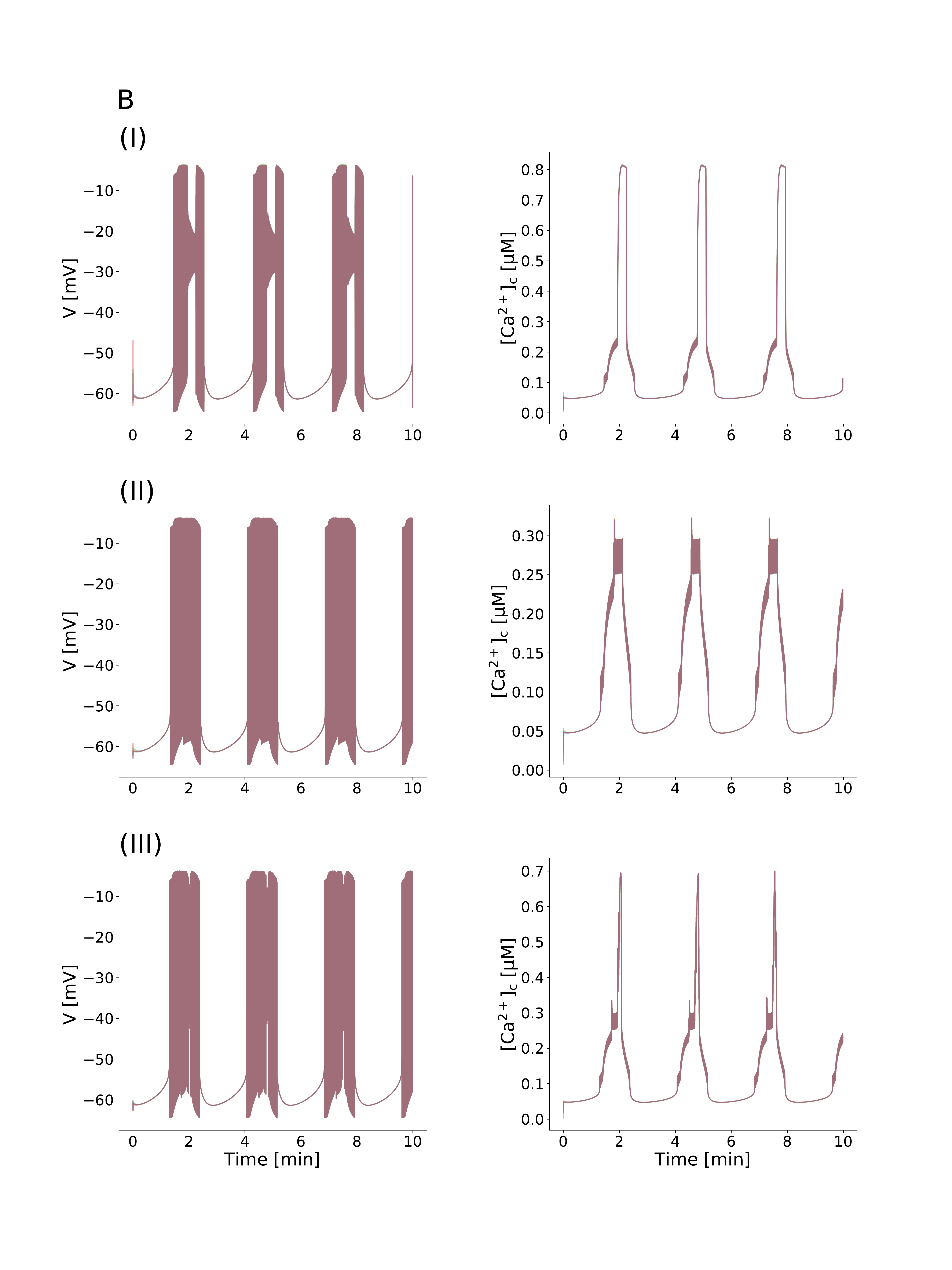}
\end{minipage}
\end{tabular}
\vspace{-1mm}
\caption{\scriptsize {\bf Electrical coupling is more effective on $\textrm{Ca}^{2+}$ activity. (A)} The interplay between electrical coupling and metabolic coupling strength in the mean $[\textrm{Ca}^{2+}]_{\textrm{c}}$ for all $\beta$-cells in a human islet model. Note that gap junction electrical coupling is more effective on intracellular $\textrm{Ca}^{2+}$ activity. {\bf(B)} Representative simulated membrane potential and $[\textrm{Ca}^{2+}]_{\textrm{c}}$ time-courses corresponding to three points {\bf(I)}, {\bf(II)}, and {\bf(III)} indicated in {\bf(A)}. Changes in electrical behavior cause varying levels of $\textrm{Ca}^{2+}$ concentration. {\bf(C)},{\bf(D)} The interplay between electrical coupling and metabolic coupling strength in the synchronization of $[\textrm{Ca}^{2+}]_{\textrm{c}}$ and membrane potential oscillations. Data in {\bf(A)}, {\bf(C)}, and {\bf(D)} represent the mean $\pm$ S.E.M. over 10 modeled islets.}
\label{fig3}
\end{figure}

In this model, we found that the mean behavior of $[\textrm{Ca}^{2+}]_{\textrm{c}}$ for single $\beta$-cell changed only slightly when fixing the parameter $g_{\textrm{c}}$ and elevating the value of $P_{\textrm{G6P.F6P}}$ (Fig.\,\ref{fig3}\,\textcolor{blue(ncs)}{A}). In other words, for any fixed $g_{\textrm{c}}$ the level of intracellular $\textrm{Ca}^{2+}$ concentration did not depend strongly on $P_{\textrm{G6P.F6P}}$. It should be noted that $\beta$-cell $[\textrm{Ca}^{2+}]_{\textrm{c}}$ was noticeably low at $P_{\textrm{G6P.F6P}} = 0$, because purely electrical coupling did not synchronize metabolic oscillations, giving rise to out-of-phase slow bursting and small amplitude $\textrm{Ca}^{2+}$ oscillations. This predicted that to increase the $\textrm{Ca}^{2+}$ concentration, there had to be metabolic diffusion between islet $\beta$-cells, even a very small $P_{\textrm{G6P.F6P}}$ value.

At fixed $P_{\textrm{G6P.F6P}} > 0$, the mean $[\textrm{Ca}^{2+}]_{\textrm{c}}$ of single $\beta$-cell significantly changed with varying value of $g_{\textrm{c}}$ such that the $\textrm{Ca}^{2+}$ concentration of an active $\beta$-cell was maximum in certain ranges of $g_{\textrm{c}}$, and was then reduced $\sim 50\%$ (Fig.\,\ref{fig3}\,\textcolor{blue(ncs)}{A}). These results demonstrate that electrical coupling has a greater effect compare with metabolic coupling on the $\beta$-cell $\textrm{Ca}^{2+}$ activity in human islets.

Fig.\,\ref{fig3}\,\textcolor{blue(ncs)}{A} displayed that changes in $\beta$-cell $[\textrm{Ca}^{2+}]_{\textrm{c}}$ was bimodal. Three points were indicated in the plot at which the average $[\textrm{Ca}^{2+}]_{\textrm{c}}$ level was significantly different. These differences in $[\textrm{Ca}^{2+}]_{\textrm{c}}$ occurred with changes in multicellular electrical behavior (Fig.\,\ref{fig3}\,\textcolor{blue(ncs)}{B}). In fact, different oscillations of the $\beta$-cell membrane potential, imposing $[\textrm{Ca}^{2+}]_{\textrm{c}}$ oscillation patterns, caused varying cytosolic calcium concentration. Furthermore, Fig.\,\ref{fig3}\,\textcolor{blue(ncs)}{C} and \textcolor{blue(ncs)}{D} depicted that electrical dynamics, especially intracellular $\textrm{Ca}^{2+}$ activity, in the islet were almost fully synchronous whereas $[\textrm{Ca}^{2+}]_{\textrm{c}}$ behavior was bimodal. Therefore, as we already mentioned, the islet gap junction coupling is mostly important for synchronization of oscillatory activity.

\subsection*{4. The interplay of gap junction electrical coupling and glucose stimulation in $[\textrm{Ca}^{2+}]_{\textrm{c}}$ activity}\hypertarget{Res4}{}

\begin{figure}[!ht]
\centering
\subfigure{
\includegraphics[width=.39\textwidth]{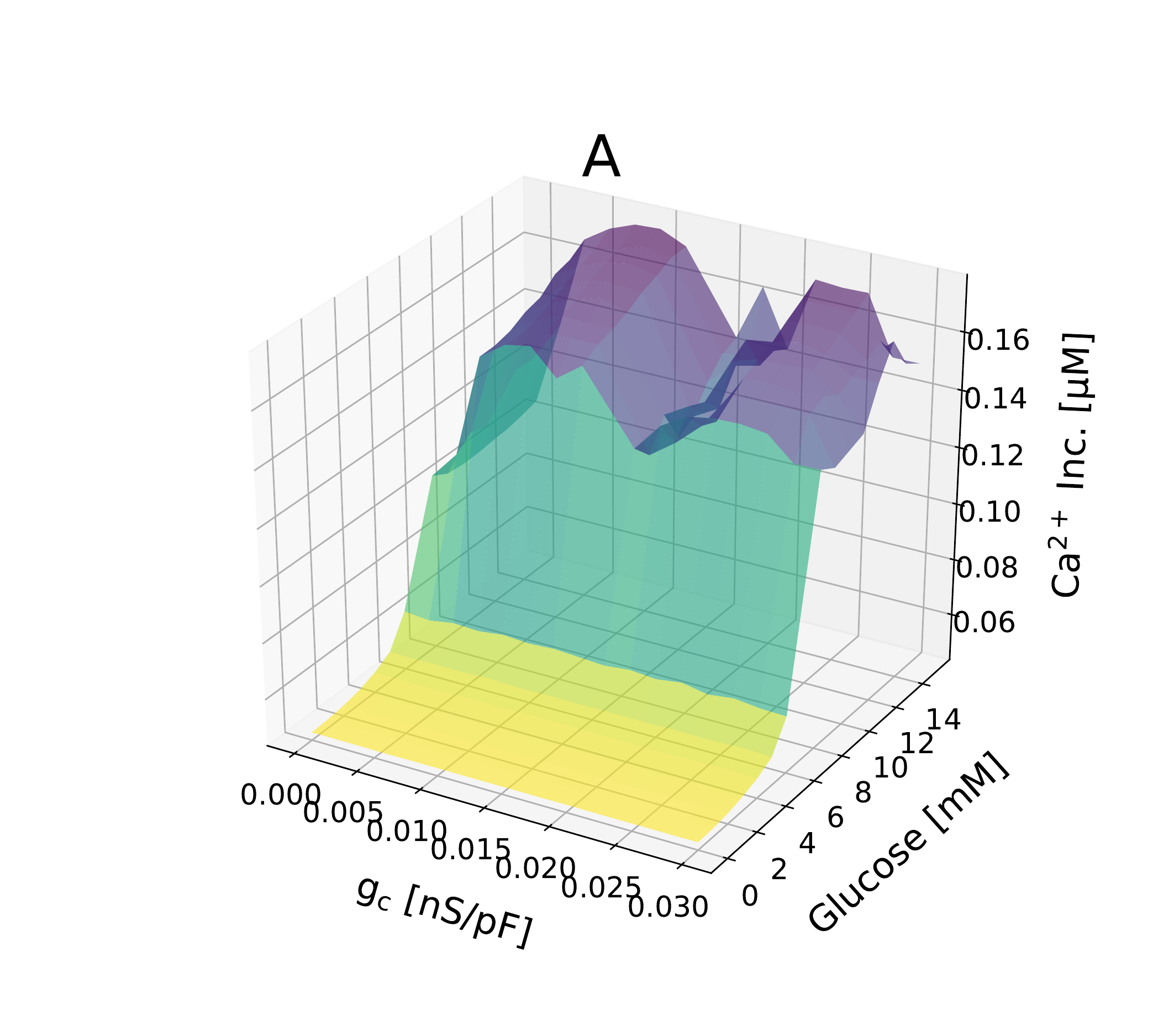}
}
\subfigure{
\includegraphics[width=.39\textwidth]{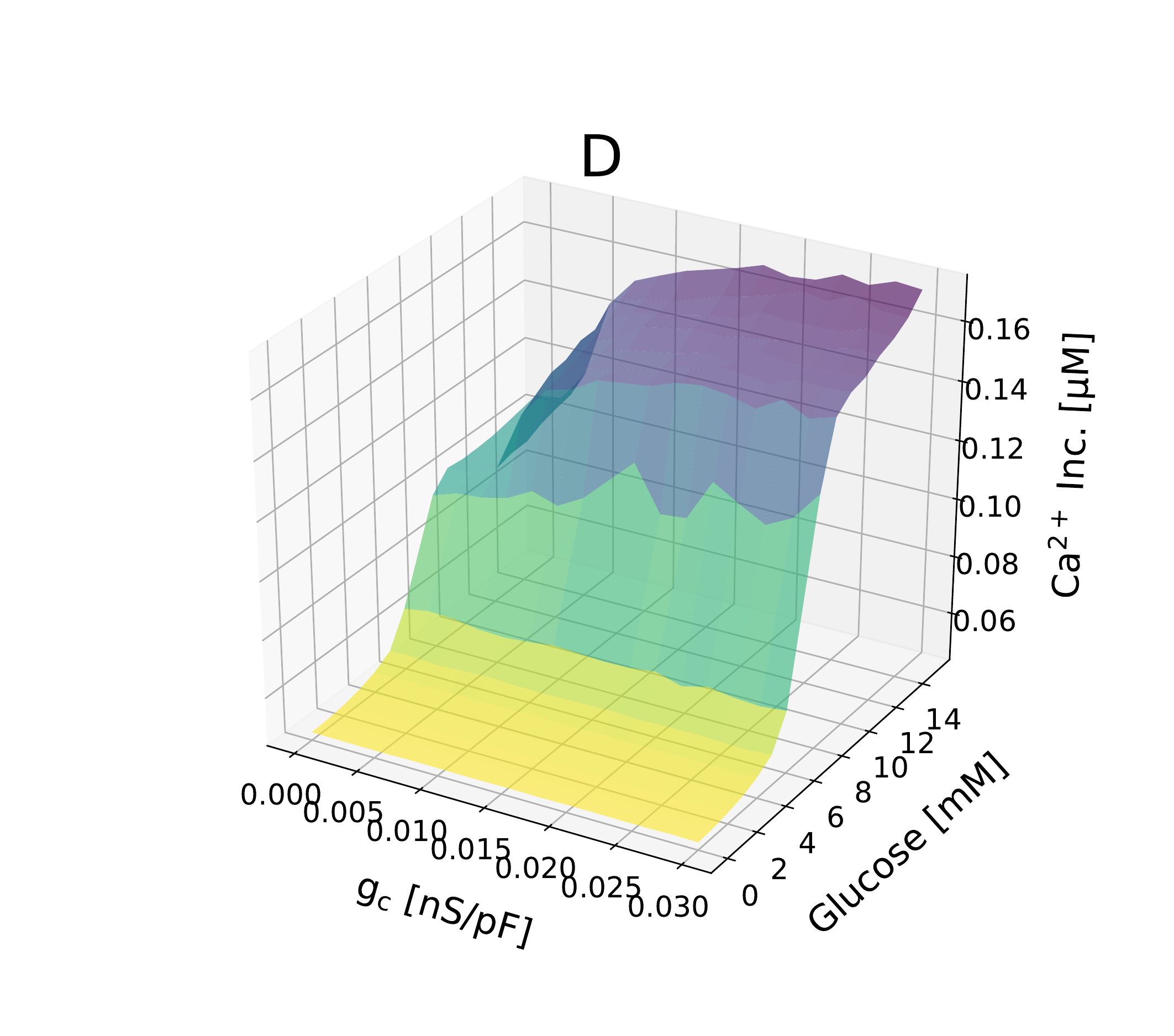}
}
\\
\subfigure{
\includegraphics[width=.39\textwidth]{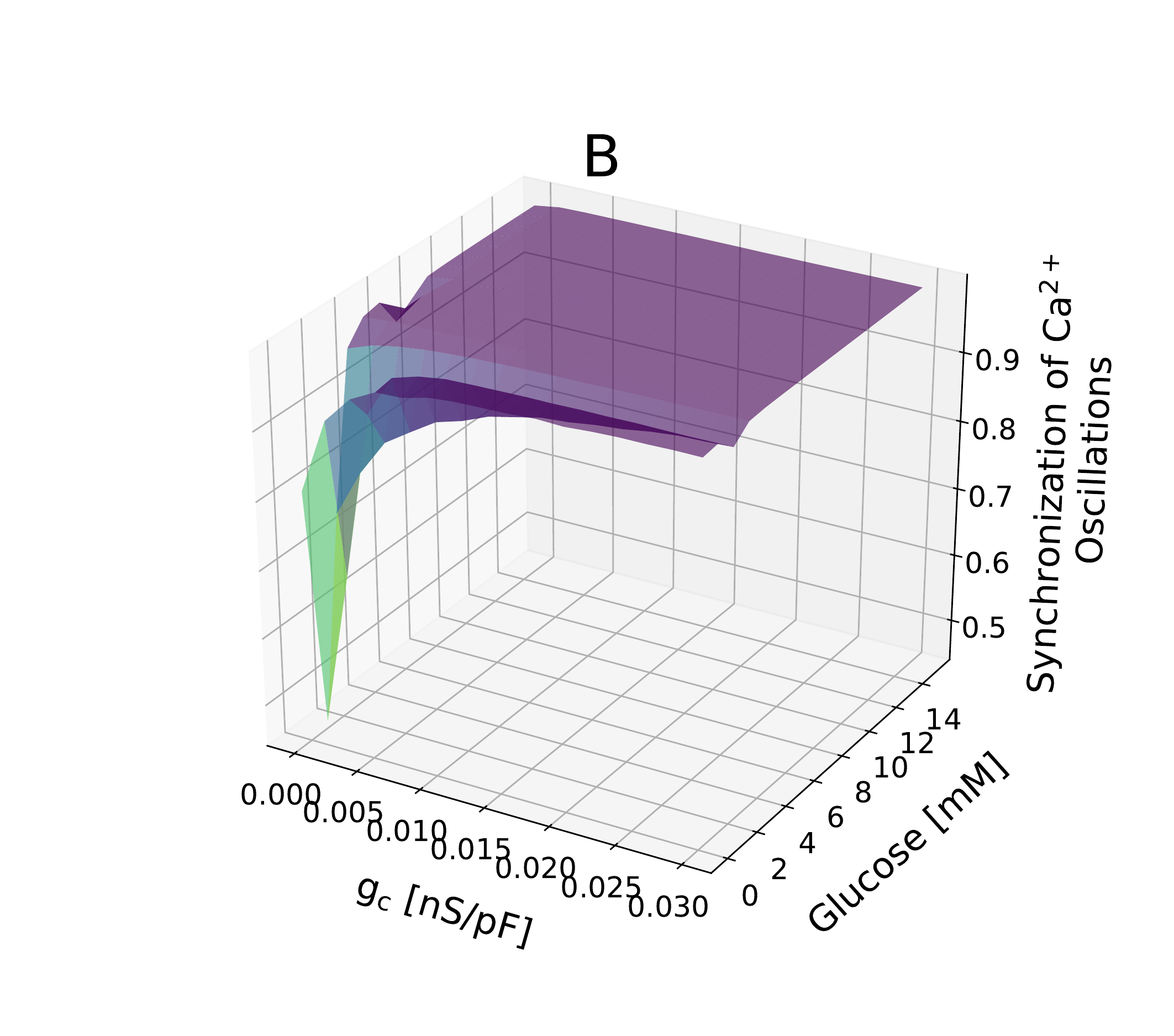}
}
\subfigure{
\includegraphics[width=.39\textwidth]{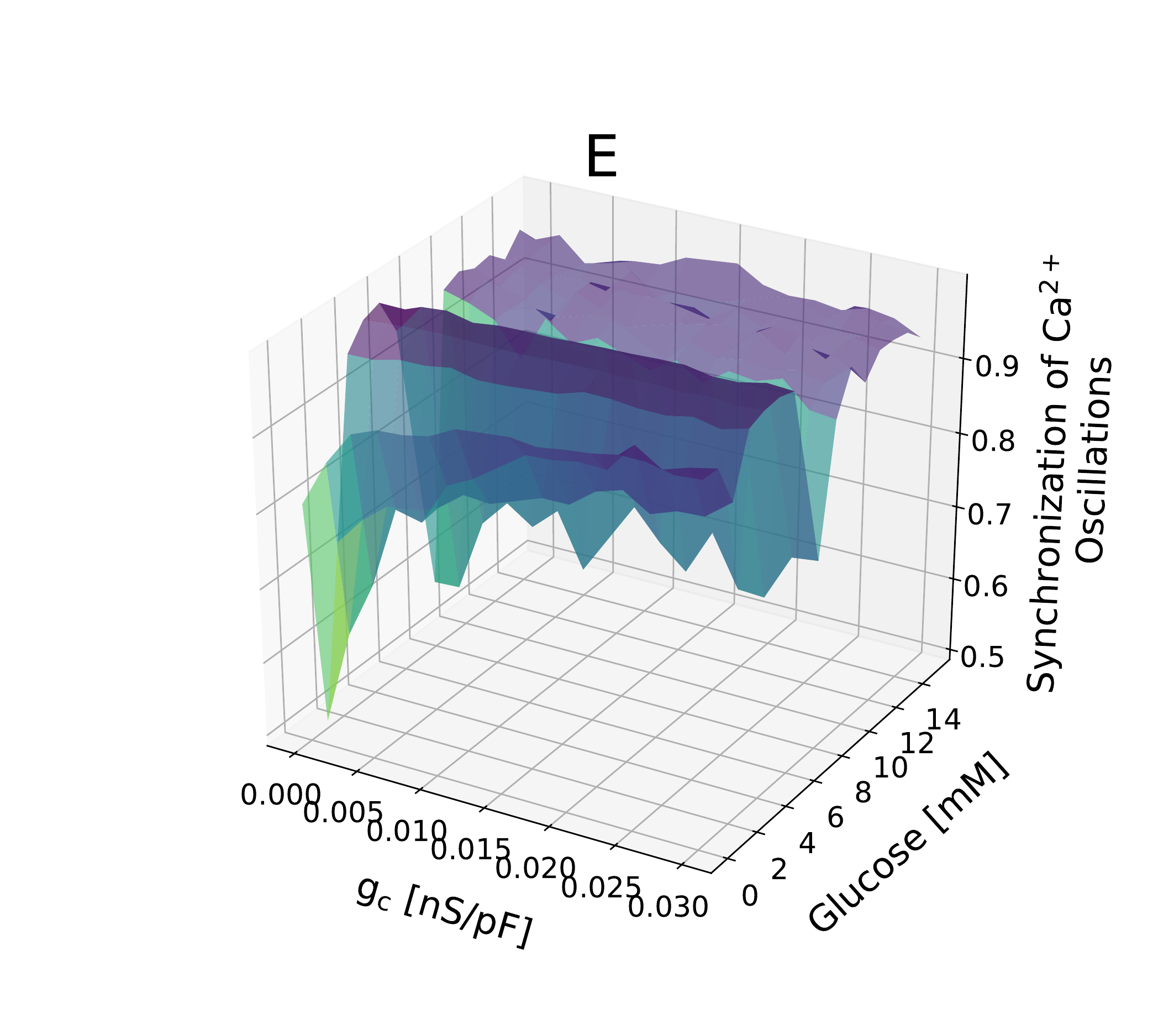}
}
\\
\subfigure{
\includegraphics[width=.39\textwidth]{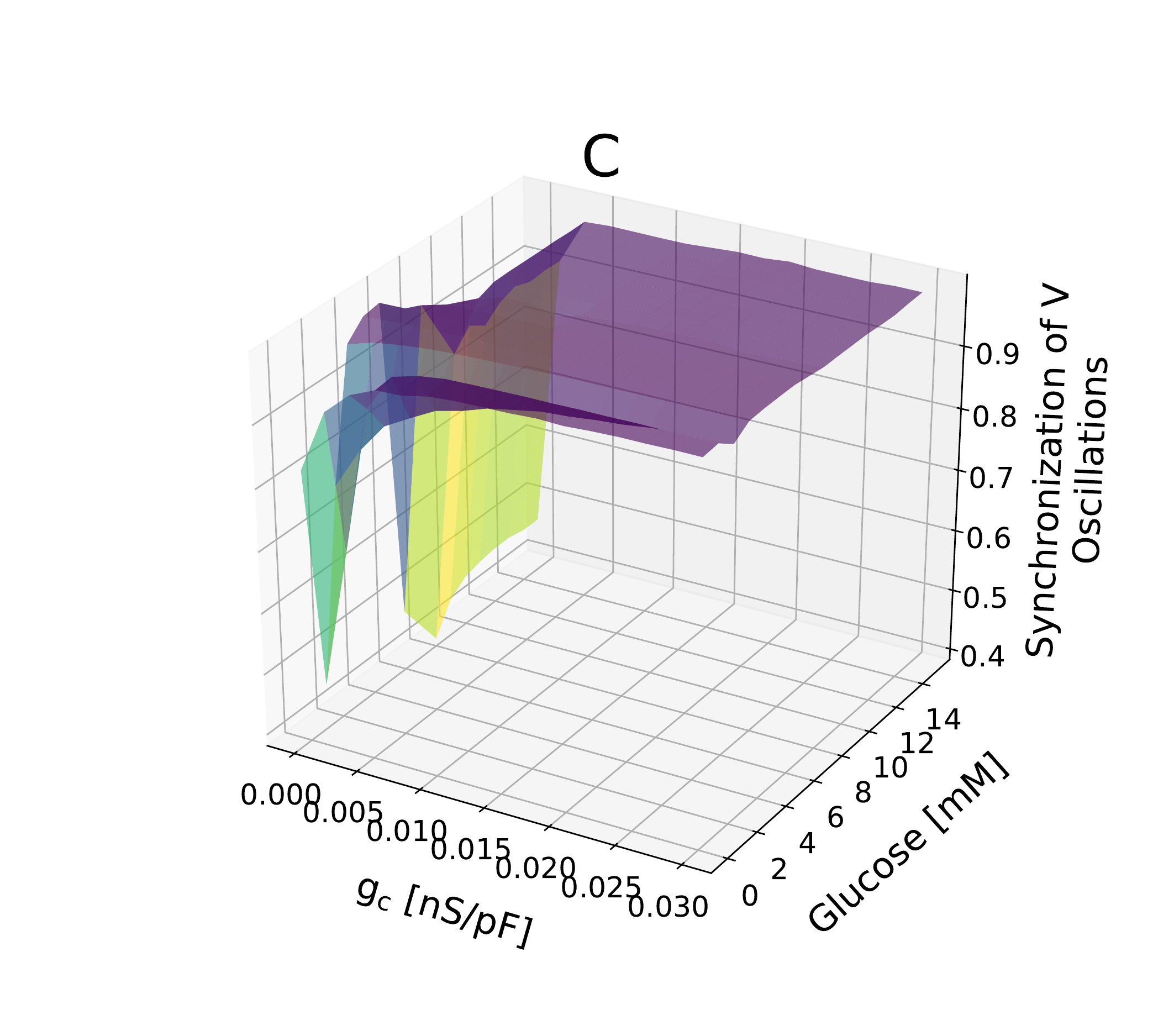}
}
\subfigure{
\includegraphics[width=.39\textwidth]{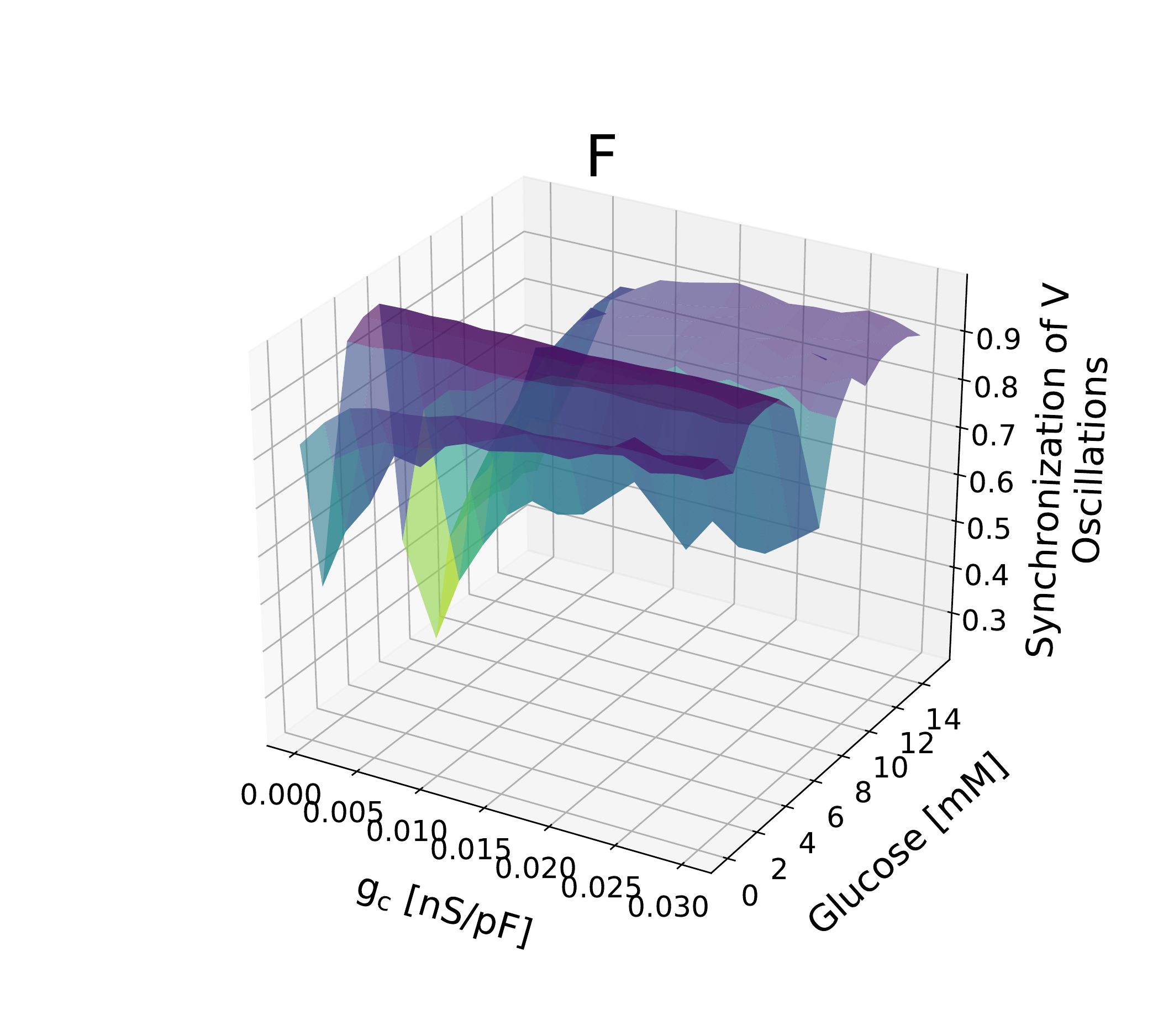}
}
\vspace{-2mm}
\caption{\scriptsize {\bf(A)},{\bf(B)},{\bf(C)} The interplay between gap junction coupling strength and glucose concentration in the mean $[\textrm{Ca}^{2+}]_{\textrm{c}}$ response, and in the synchronization of $[\textrm{Ca}^{2+}]_{\textrm{c}}$ and membrane potential oscillations across the intact islet. {\bf(D)},{\bf(E)},{\bf(F)} As in panels {\bf(A)},{\bf(B)},{\bf(C)}, but across the islet lacking $\sim 55\%$ $\beta$-cells. Note that after loss of $\sim 55\%$ $\beta$-cell mass the level of intracellular $\textrm{Ca}^{2+}$ is more slowly saturated, and the islet synchronization is more sensitive to increasing glucose level. Data represent the mean $\pm$ S.E.M. over 10 modeled islets.}
\label{fig4}
\end{figure}

We examined the impact of islet gap junction coupling, and elevated glucose concentrations on the dynamics of intracellular $\textrm{Ca}^{2+}$. At low levels of glucose $(< 5 \, \textrm{mM})$, despite increasing intercellular coupling strength and nearly complete synchrony of electrical activity, $\beta$-cells showed globally quiescent behavior, and were inexcitable within the islet (Fig.\,\ref{fig4}\,\textcolor{blue(ncs)}{A} and \textcolor{blue(ncs)}{B} and \textcolor{blue(ncs)}{C}). In other words, the coupling strength between human $\beta$-cells had no considerable affect on $\textrm{Ca}^{2+}$ concentration of single $\beta$-cell in basal levels of glucose. At $5-10\,\textrm{mM}$ glucose, the islet $\beta$-cells sharply transferred from inactive to active states, and glucose acutely elevated $[\textrm{Ca}^{2+}]_{\textrm{c}}$. In fact, a phase transition between globally active behavior and global quiescence of $\beta$-cells was observed, as cellular excitability approached a critical threshold. For high levels of glucose $(> 10\,\textrm{mM})$, the slope of $[\textrm{Ca}^{2+}]_{\textrm{c}}$ elevation significantly decreased, and intracellular $\textrm{Ca}^{2+}$ activity comparatively achieved saturation. Indeed, in the specific range $5-10\,\textrm{mM}$, the $\beta$-cell $\textrm{Ca}^{2+}$ dynamics and concentrations were highly sensitive to increasing glucose level and then $[\textrm{Ca}^{2+}]_{\textrm{c}}$ was only slightly altered, leading to near-independent insulin behavior of the glucose gradient. Also, at high levels of glucose, we observed bimodal behavior in cytosolic calcium concentrations, similar to Fig.\,\ref{fig3}, caused by varying values of $g_{\textrm{c}}$ and very different oscillation patterns. The critical point $(\textrm{e.g.} \sim 5\,\textrm{mM})$ did not depend on the value of gap junction conductance, whereas the strength of coupling could affect the behaviour above the threshold point such that in small values of $g_{\textrm{c}}$, intracellular calcium concentration was saturated at higher glucose levels. Fig.\,\ref{fig4}\,\textcolor{blue(ncs)}{B} and \textcolor{blue(ncs)}{C} showed that interestingly, change in glucose levels had no significant effect on global synchronization of electrical activity across the islet.

Additionally, we eliminated $\sim 55\%$ the islet $\beta$-cells, and then analyzed the role of $g_{\textrm{c}}$ and glucose levels on $[\textrm{Ca}^{2+}]_{\textrm{c}}$  activity (Fig.\,\ref{fig4}\,\textcolor{blue(ncs)}{D} and \textcolor{blue(ncs)}{E} and \textcolor{blue(ncs)}{F}). As above, intracellular $\textrm{Ca}^{2+}$ concentrations of single $\beta$-cell and the release of insulin exhibited a steep sigmoidal secretory response to increasing glucose levels. The transition between quiet state and avalanche occurred at a position equivalent to $\sim 5 \, \textrm{mM}$ glucose, similar to intact islet, while at $5-10 \, \textrm{mM}$ glucose, the slope of $[\textrm{Ca}^{2+}]_{\textrm{c}}$ elevation was less in islet lacking $\sim 55\%$ $\beta$-cells compared with the intact islet, i.e., the level of intracellular $\textrm{Ca}^{2+}$ was more slowly saturated. During high glucose concentrations, $[\textrm{Ca}^{2+}]_{\textrm{c}}$ in islets lacking $\beta$-cells was relatively unchanged from the intact islet, although in small values of $g_{\textrm{c}}$ there were noticeable differences (Fig.\,\ref{fig4}\,\textcolor{blue(ncs)}{D}). Furthermore, when the level of glucose concentrations was elevated, islet synchrony of electrical patterns had complex behavior. In fact, the synchronization of islet lacking $\sim 55\%$ $\beta$-cells was less than intact islet and more sensitive to increasing glucose level (Fig.\,\ref{fig4}\,\textcolor{blue(ncs)}{E} and \textcolor{blue(ncs)}{F}).

\subsection*{5. Kv-channels increase $[\textrm{Ca}^{2+}]_{\textrm{c}}$ levels in the absence of gap junction coupling}\hypertarget{Res5}{}

\begin{figure}[!ht]
\centering
\vspace{-6mm}
\subfigure{
\includegraphics[width=5.2cm]{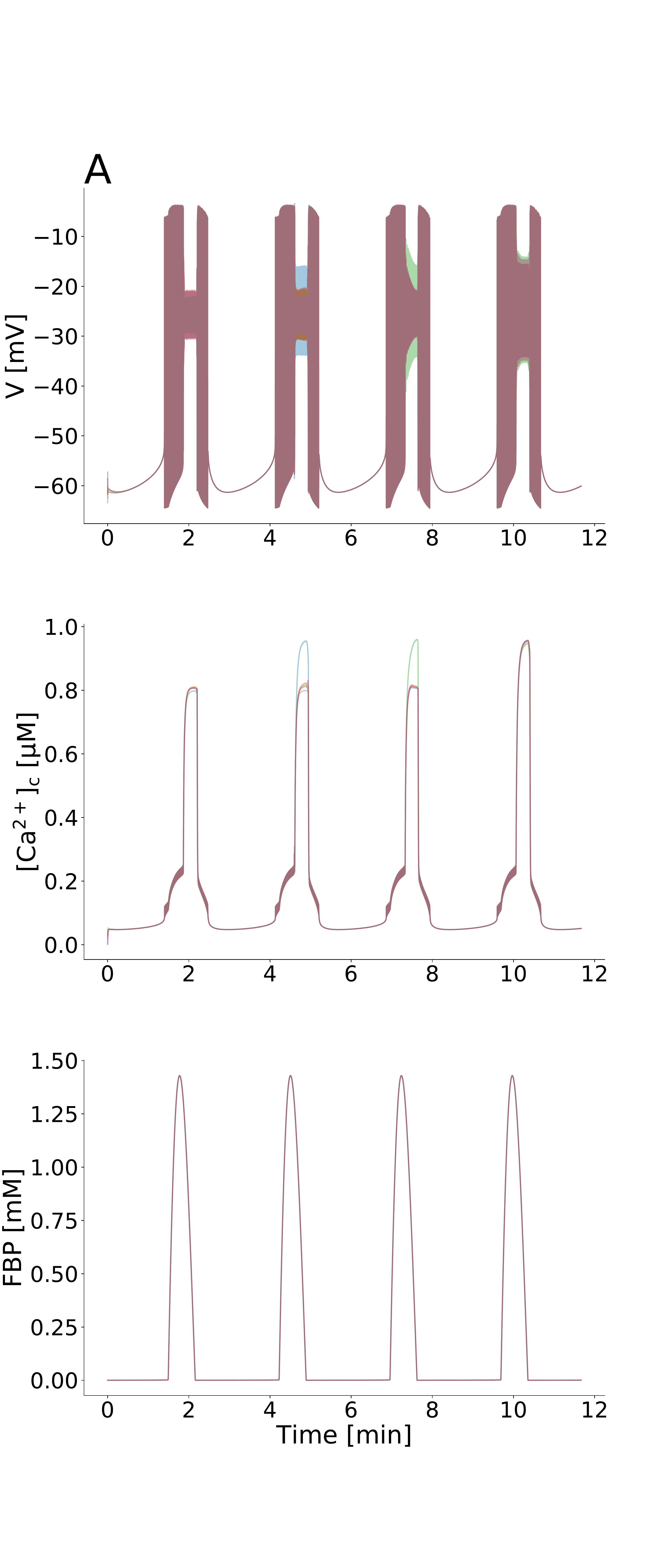}
}
\subfigure{
\includegraphics[width=5.2cm]{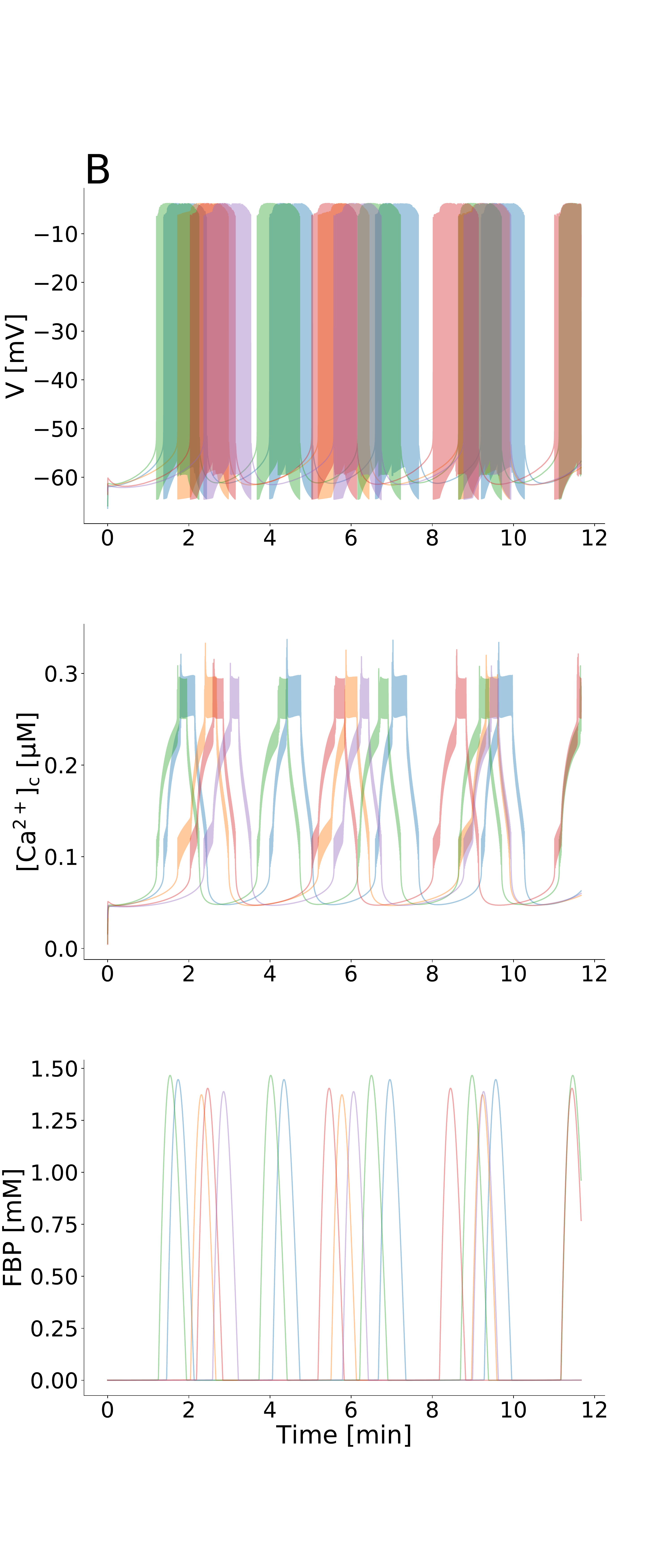}
}
\subfigure{
\includegraphics[width=5.2cm]{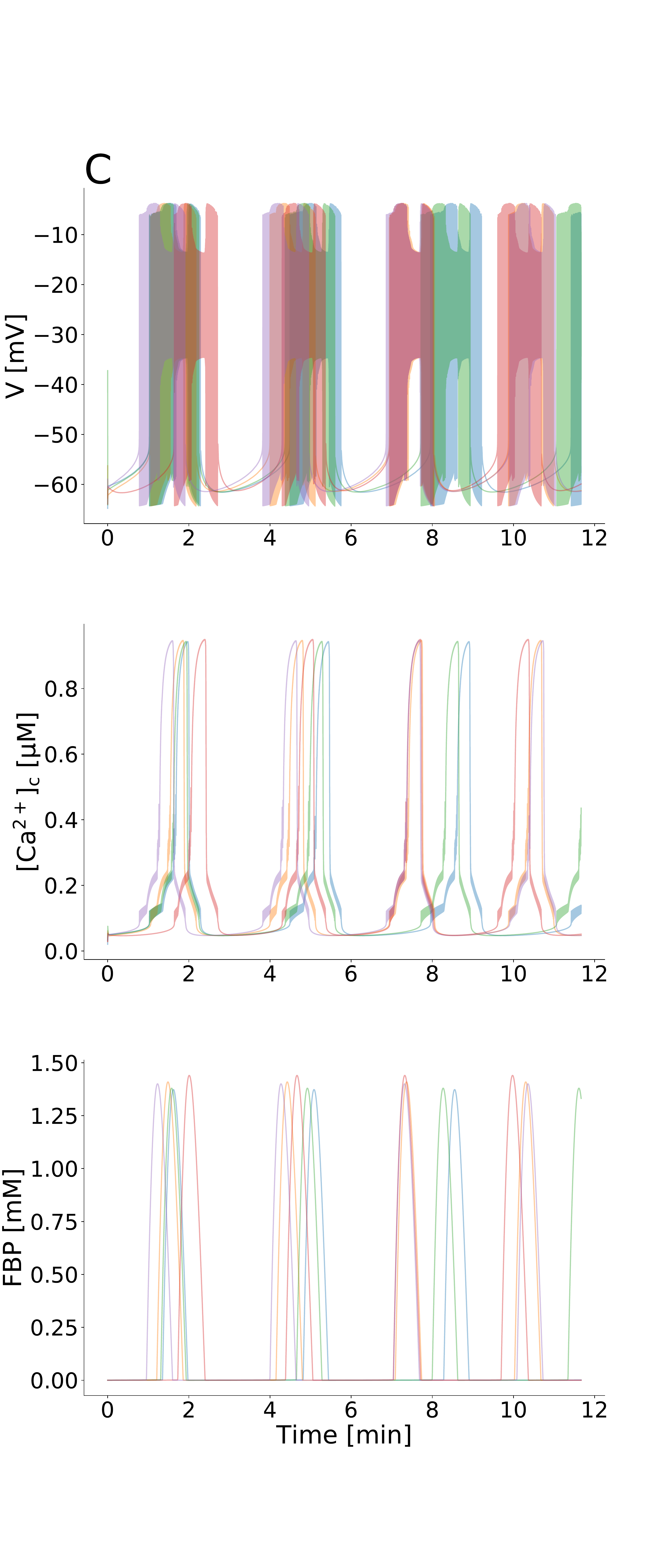}
}
\vspace{-8mm}
\caption{\scriptsize {\bf Effect of Kv-channels on $[\textrm{Ca}^{2+}]_{\textrm{c}}$ patterns. (A)} Membrane potential, cytosolic calcium, and FBP time series for coupled human $\beta$-cells with $g_{\textrm{Kv}} = 0.215\,\,\textrm{nS}\,\,\textrm{pF}^{-1}$. {\bf(B)} As in panel {\bf(A)}, but for uncoupled $\beta$-cells. {\bf(C)} Model simulation of uncoupled $\beta$-cells as in panel {\bf(B)}, but with $g_{\textrm{Kv}} = 0.200\,\,\textrm{nS}\,\,\textrm{pF}^{-1}$.}
\label{fig5}
\end{figure}

We predict that Kv-channels modify the electrical patterns of $\beta$-cells lacking gap junctions and that changes in $\beta$-cell burst behavior cause quantitatively alterations in the intracellular $\textrm{Ca}^{2+}$ concentration. To test this prediction, we first considered the simulated behavior of interconnected $\beta$-cells, which had Kv-conductances picked from a normal distribution (Fig.\,\ref{fig5}\,\textcolor{blue(ncs)}{A}), and then determined how different $g_{\textrm{Kv}}$ values affected the slow electrical burst patterns. When the modeled $\beta$-cells were isolated from the islet, not only the stochastic and heterogeneous behavior occurred but the shape of the oscillatory pattern of the membrane potential and $[\textrm{Ca}^{2+}]_{\textrm{c}}$ changed, which for the amplitude of $\textrm{Ca}^{2+}$ oscillations was much lower compared to the coupled scenario (Fig.\,\ref{fig5}\,\textcolor{blue(ncs)}{B}). In continuation, we decreased slightly Kv-channel conductance values of dissociated $\beta$-cells, and compared activity patterns of these cells with the coupled $\beta$-cell behavior. Surprisingly, the electrical patterns, specifically $\textrm{Ca}^{2+}$ activity and the metabolic oscillations resulting from a variation of conductance $g_{\textrm{Kv}}$ nicely resembled the global islet behavior mediated through gap junctional communications and synchronizing dynamics, as summarized in Fig.\,\ref{fig5}\,\textcolor{blue(ncs)}{A} and \textcolor{blue(ncs)}{C}. These findings suggest that modifying forms of electrical activity in human $\beta$-cells to enhance insulin levels can arise out of another regulatory mechanism without coupling and synchrony across the islet, i.e., small changes in the expression of Kv-channels lead to unexpected changes in the integrated behavior of all $\beta$-cells. In sum, according to this view changes in oscillatory behavior of $\beta$-cells involving different mechanisms play a physiologically important role in the level of $[\textrm{Ca}^{2+}]_{\textrm{c}}$ and insulin secretion.

\section*{Discussion}\hypertarget{Dis}{}

Type 2 diabetes mellitus (T2DM) is a complex metabolic disorder triggered by a progressive decline in $\beta$-cell mass and an increase in insulin resistance. In T2D, there are fundamental changes in insulin secretion dynamics, which can lead to the development of insulin insufficiency. The total amount of released insulin is predominantly the product of the morphological mass of $\beta$-cells and insulin output of each of these cells. Thus, inadequate levels of insulin are importantly the result of deficiency in either $\beta$-cell mass or function, or both, leading to hyperglycemia and diabetes. Currently, diabetes research shows that the actual cause of developing T2D seems to be strongly correlated with pancreatic $\beta$-cells; however it has long been assumed that insulin resistance is the major risk component. In patients with type 2 diabetes, apoptotic $\beta$-cell death can be induced by various etiological factors, such as exposure to chronic hyperglycemia and carbohydrate metabolites \cite{90}, oxidative stress \cite{97,96}, pro-inflammatory mediators (e.g., cytokines) \cite{59}, proinsulin misfolding \cite{60,58}, human islet amyloid polypeptide misfolding \cite{61}, as well as endoplasmic reticulum stress \cite{62}, which these sequential events initiate apoptosis, increase $\beta$-cell workload and stress, culminate in exhaustion, and finally $\beta$-cell death \cite{66,63,64,65,67}. Therefore, creating a successful treatment for T2D will need to specifically include targeting insulin resistance, regenerating $\beta$-cell mass, and restoring appropriate insulin release by recovery and increase of $\beta$-cell function. Despite these findings, the precise relationship between morphological and functional $\beta$-cell in progressing impairments in insulin secretion is a thorny issue, and dynamics of $\beta$-cell dysfunction and $\beta$-cell loss in human T2D phenotype is still under debate \cite{46,47,48}.

To get a more detailed insight into the distinct role of human $\beta$-cell mass and function as cause for type 2 diabetes, we made use of multicellular computational approach of interconnected $\beta$-cells, based on the theoretical model of Riz et al. \cite{50}. To incorporate known particularities, $\beta$-cell heterogeneity was characterized by picking parameter values from normal distributions, and $\beta$-cell mass reduction was established in the islet network by removing simulated cells, randomly. We found that contribution of impaired $\beta$-cell function to insulin inadequacy, correlated with irregular $\textrm{Ca}^{2+}$ activity dynamics, could be deduced from $\beta$-cell mass deficit in type 2 diabetes (Fig.\,\ref{fig1}). Besides alterations in the summed $[\textrm{Ca}^{2+}]_{\textrm{c}}$ activity after removal of insulin secreting cells (Fig.\,\ref{fig1}\,\textcolor{blue(ncs)}{A}), surprisingly the human islet showed decreased levels in the average $[\textrm{Ca}^{2+}]_{\textrm{c}}$ of an active $\beta$-cell (Fig.\,\ref{fig1}\,\textcolor{blue(ncs)}{B}), caused by reduced $\textrm{Ca}^{2+}$ concentration uptake and asynchronous $[\textrm{Ca}^{2+}]_{\textrm{c}}$ oscillations (Fig.\,\ref{fig1}\,\textcolor{blue(ncs)}{E} (I and II)). Our results indicate that in addition to disruption in $\beta$-cell population electrical activity, changes in the form of oscillatory patterns of membrane potential and intracellular $\textrm{Ca}^{2+}$ concentration evoke functional deteriorations in $\beta$-cells, such as insulin secretory dysfunction, which are caused by changes in $\beta$-cell mass. However, determining the specific role of $\beta$-cell mass on $\beta$-cell function is beyond the scope of this manuscript. Therefore, future studies will be needed to precisely examine whether a decrease in $\beta$-cell mass primarily is the leading factor in $\beta$-cell dysfunction.

Pulsatile insulin secretion is tightly associated with the synchronous oscillations in electrical activity, particular oscillatory dynamics of $[\textrm{Ca}^{2+}]_{\textrm{c}}$ within the islet. Insulin pulsatility leads to complex rhythms in blood glucose concentration \cite{69,30}, enhances hepatic insulin action and post-receptor signaling \cite{70}, protects against insulin resistance and shows a greater efficiency than continuous levels of secretion \cite{33}. Previous experimental investigations have revealed that a decline in first-phase insulin response and an impairment in regular rhythmic secretion patterns can be found at the onset of T2D and thereafter \cite{33,37}, which occurred in $\sim 50\%$ deficit of $\beta$-cell mass \cite{43,42,45}. Our computational modeling confirms these experimental observations, as $\sim 50\%$ of $\beta$-cells are lost, the synchronous membrane potential and $[\textrm{Ca}^{2+}]_{\textrm{c}}$ excursions, underlying the peak amplitude of early-phase insulin release and the coordination of insulin pulsatility fashion, began to be disrupted within the islet cellular network (Fig.\,\ref{fig1}\,\textcolor{blue(ncs)}{C} and \textcolor{blue(ncs)}{D}).

Cx36 gap junctions have potential roles in dynamics and physiological function of the islet \cite{16,20}, and a reduction in $\beta$-cell coupling has been suggested to occur in type 2 diabetes \cite{14,38,40,71}. Inter-$\beta$-cellular coupling has also been implicated in protecting $\beta$-cells against a variety of cytotoxic factors, regulating $\beta$-cell differentiation and maturation, and supporting islet development and fitness \cite{68,72,73}. Chronic hyperglycemia characterizing the onset of diabetes leads to impaired gap junctional communication \cite{14}, which makes the islets more sensitive to $\beta$-cell death \cite{34,41}. Likewise, it has been demonstrated that a number of chronic insults, including glucotoxicity \cite{74}, lipotoxicity \cite{89,21}, and pro-inflammatory cytokines and oxidative stress \cite{75} target the expression of Cx36 transcript and inhibit gap junction functionality. Therefore, human $\beta$-cell islets exposed to such insults exhibit a lack in intra-islet synchrony of $[\textrm{Ca}^{2+}]_{\textrm{c}}$ oscillations, a suppression and limited propagation of calcium waves, a disruption in plasma insulin pulsatility, and glucose intolerance \cite{78,79,80}. A significant decrease in the peak level of first-phase insulin release and loss of pulsatile second-phase secretion is mainly the secretory defects observed in human T2D, resulting in disrupted glucose homeostasis \cite{37}. For the most part, the peak elevation of first-phase secretion and the second-phase pulses are dependent on the coordinated pulsatility of individual islets and eventually abolish as the development of diabetes. Furthermore, cell-cell interactions via gap junctional channels are a prerequisite for the oscillatory patterns of electrical activity within the islet and regulate the dynamics of insulin secretion.

Our analysis regarding the interplay between loss of gap junction coupling and impairment of synchronous electrical activity patterns in inadequate levels of plasma insulin and progression of T2D suggest a essentially important impression for intercellular connections. In fact, firstly the level of $\textrm{Ca}^{2+}$ concentration immediately decreased following reduction in $\beta$-cell coupling, however the oscillations of $\beta$-cell membrane potential and $[\textrm{Ca}^{2+}]_{\textrm{c}}$ synchronized (Fig.\,\ref{fig2}). Secondly, the $\beta$-cell population in islets showed poorly coordinated behavior, especially $[\textrm{Ca}^{2+}]_{\textrm{c}}$ dynamics, after $\sim 50 \%$ $\beta$-cell coupling loss, although was harmonize as $< 50 \%$ (Fig.\,\ref{fig2}\,\textcolor{blue(ncs)}{B} and \textcolor{blue(ncs)}{C}). The data discussed above strongly supports that the role of gap junctional coupling in affecting the cytosolic calcium concentration and the amount of secreted insulin is more subtle and fundamental than synchronous oscillations of $\beta$-cell activity.

Available studies suggest that mouse and human $\beta$-cells show different electrical dynamics; mouse $\beta$-cell population display islet-wide synchrony in response to glucose, whereas human islet synchrony of $\textrm{Ca}^{2+}$ oscillations is constrained to localized subpopulations \cite{98,99}. These differences probably relate to differences in mouse and human islet architectures: mouse islets have a large, highly connected $\beta$-cell core, whereas human islets are composed of distinct clusters of gap junction coupled $\beta$-cells \cite{99,56,100}. Noteworthy, in the case of intra-islet synchronization, our results seem to be in contrast to less coordinated behavior in human islets. These findings highlight that $\beta$-cells in human islets occur in distinct clusters separated by other cell types, notably $\alpha$-cells and vascular cells \cite{99}.

Using computer simulations, we proposed that human $\beta$-cells exhibit great different behavior in $\textrm{Ca}^{2+}$ dynamics caused by a substantial decrease in islet gap junction coupling, i.e., lack of
$> 50 \%$ $\beta$-cell connections sped up reducing $[\textrm{Ca}^{2+}]_{\textrm{c}}$ level in single $\beta$-cell (Fig.\,\ref{fig2}\,\textcolor{blue(ncs)}{A}) and lost inter-$\beta$-cell synchronization (Fig.\,\ref{fig2}\,\textcolor{blue(ncs)}{B} and \textcolor{blue(ncs)}{C}), which is believed to lead to impairment of calcium waves and normal oscillatory insulin secretion \cite{27}. In particular, our results show $\beta$-cell functional insufficiency, such that there exist specific changes in the oscillation patterns of $\beta$-cell electrical activity (not shown), which most likely be due to combined increased $\beta$-cell apoptosis and workload, and finally result in functional exhaustion and persistent hyperglycemia.

Gap junctional coupling between $\beta$-cells provide the intra-islet synchrony of glycolysis oscillations, which is a prerequisite for pulsatile insulin secretion \cite{107}. Slow oscillations in metabolism of glucose-stimulated $\beta$-cells coupled to electrical activity patterns by oscillations in ATP production and closure of $\textrm{K}_{\textrm{ATP}}$ channels are mediated by the positive feedback on the allosteric enzyme phosphofructokinase (PFK) via its product fructose-1,6-bisphosphate (FBP). A rise in substrate G6P, which is converted to F6P, from glucose consumption leads to FBP production, which increases the autocatalytic activation of the enzyme PFK with an eventual crash in the FBP level due to depletion of substrate G6P \cite{81}. Gap junctional permeability leads to diffusion of G6P among islet $\beta$-cells, which is considerably smaller than other glycolytic metabolites \cite{28}.

In order to investigate the effects of coupling electrically and metabolically on the oscillatory behavior of human pancreatic $\beta$-cells, we used dynamical mathematical model incorporating glycolytic oscillations. Diffusive coupling of metabolites, specifically at low $g_{\textrm{c}}$, made an important contribution to the normal function of $\beta$-cells, however significance of this diffusion process is likely less compared to electrical conduction in regulating islet activity. Namely, metabolic communications, if did not exist, could have a negative effect on the $[\textrm{Ca}^{2+}]_{\textrm{c}}$ levels (Fig.\,\ref{fig3}\,\textcolor{blue(ncs)}{A}) and a desynchronizing effect on the bursting activity as well as $\textrm{Ca}^{2+}$ dynamics of system coupled electrically through gap junctions (Fig.\,\ref{fig3}\,\textcolor{blue(ncs)}{C} and \textcolor{blue(ncs)}{D}), nevertheless increasing electrical coupling strength further was able to overcome islet asynchronous. Addition of a very small degree of metabolic coupling could contribute to intracellular $\textrm{Ca}^{2+}$ elevation and robust coordination of membrane potential and $[\textrm{Ca}^{2+}]_{\textrm{c}}$ oscillations. On the other hand, our results revealed different modes of $\beta$-cell behavior, strongly depending on the relative strength of electrical coupling. Incorporation of enhanced electrical coupling in addition to gap junctional diffusion persuaded great alterations in the intercellular calcium levels (Fig.\,\ref{fig3}\,\textcolor{blue(ncs)}{A}), and more importantly imposed the bimodal behavior on the single $\beta$-cell $[\textrm{Ca}^{2+}]_{\textrm{c}}$ achieved through widely transforming patterns of membrane potential and calcium oscillations (Fig.\,\ref{fig3}\,\textcolor{blue(ncs)}{B}).

$\beta$-cells within islets of Langerhans respond to stimulatory glucose level by insulin secretion. Electrical coupling across the islets, principally via Cx36 gap junction channels,  mediates oscillatory dynamics of membrane depolarization and $[\textrm{Ca}^{2+}]_{\textrm{c}}$ to propagate $\textrm{Ca}^{2+}$ waves and robust pulsatile insulin release under elevated glucose, as well as efficient suppression of spontaneous $[\textrm{Ca}^{2+}]_{\textrm{c}}$ elevations under basal glucose \cite{87,22}, illustrating a strong link between glucose-stimulated insulin secretion and gap junction function. Upon a glucose gradient, a characteristic sigmoidal secretory response is observed in intact islets, indicating critical behavior that depends on physiological properties of gap junction conductance. Additionally, intact islets exhibit more insulin response to increasing glucose than dispersed $\beta$-cells and that, at nonstimulatory glucose concentrations, insulin levels from dispersed $\beta$-cells are significantly higher than from intact islets \cite{34}. It is therefore important to consider cellular communication for regulating insulin secretory dynamics and ultimately glucose homeostasis. The disruption of gap junctional coupling results in reduced first-phase amplitude of insulin secretion, loss of coordinated $[\textrm{Ca}^{2+}]_{\textrm{c}}$ oscillations leading to lack of pulsatile second-phase insulin release, and disrupted glucose homeostasis \cite{22,34,41}, similar to defects seen in human patients with type 2 diabetes \cite{33,37,40}. More striking is the fact that islets lacking gap junctions have statistically normal insulin levels and insulin sensitivity, despite glucose intolerance due to altered dynamics of insulin secretion.

We simulated $\beta$-cell behavior under varying the glucose concentration periodically from 0 to 14 mM to determine the relative role of glucose levels and gap junction activity in shaping the glucose-stimulated $[\textrm{Ca}^{2+}]_{\textrm{c}}$ of an active $\beta$-cell and its ability to insulin secretion. In clusters with $\sim 55 \%$ $\beta$-cell loss, $[\textrm{Ca}^{2+}]_{\textrm{c}}$ responses after glucose stimulation were characterized by a sharp transition phase between quiescent and active behavior, and plateau phase that followed, similar to what was measured in intact islets (Fig.\,\ref{fig4}\,\textcolor{blue(ncs)}{A} and \textcolor{blue(ncs)}{D}). Most importantly, our results showed that gap junction coupling strength did not significantly impact the level of intracellular $\textrm{Ca}^{2+}$ concentration at basal glucose and the position of activation threshold, while the plateau fraction of $[\textrm{Ca}^{2+}]_{\textrm{c}}$ elevation was gap junctional dependent, later  saturating at low coupling conductances. In addition, the synchrony of electrical dynamics across the islet seemed to be almost independent on glucose levels, in case of $0 \%$ loss (Fig.\,\ref{fig4}\,\textcolor{blue(ncs)}{B} and \textcolor{blue(ncs)}{C}) and $\sim 55 \%$ loss (Fig.\,\ref{fig4}\,\textcolor{blue(ncs)}{E} and \textcolor{blue(ncs)}{F}) with increasing the coupling strength.

In the islet, the shape of electrical behavior is highly dependent on the biophysical characterizations of ion channels expressed in a single $\beta$-cell. Variability in the gating dynamics of specific channels between $\beta$-cells leads to generate variable patterns of membrane potential and $[\textrm{Ca}^{2+}]_{\textrm{c}}$ oscillations. Pedersen has demonstrated the capability of Kv-channels to change spiking behaviors to bursting patterns in human $\beta$-cells \cite{49}. Riz et al. \cite{101} and Montefusco et al. \cite{102} also investigated the contribution of $\textrm{K}^{+}$ channels in shaping $\beta$-cell electrical activity and controlling insulin secretion.

Our simulation data revealed that human $\beta$-cells lacking intercellular coupling exhibit similar electrical patterns to coupled $\beta$-cells within the islet by smoothly changing the conductance Kv-channel gating, which significantly affected the level of intracellular $\textrm{Ca}^{2+}$ concentrations. The multicellular behavior of the islet was analyzed, based on the absence of gap junctional connections, to quantitatively describe changes in $\beta$-cell $\textrm{Ca}^{2+}$ dynamics after small changes in the expression of delayed rectifying potassium channels. When the Kv-channel conductance was slowly reduced, the shape of electrical activity and oscillatory $[\textrm{Ca}^{2+}]_{\textrm{c}}$ introduced by low amplitude excursions in Fig.\,\ref{fig5}\,\textcolor{blue(ncs)}{B} modified, which were similarly observed before disruptions to gap junction coupling in intact islet (Fig.\,\ref{fig5}\,\textcolor{blue(ncs)}{A} and \textcolor{blue(ncs)}{C}). It appears that the spatiotemporal organization of $[\textrm{Ca}^{2+}]_{\textrm{c}}$ response are likely governed by two different mechanisms, characterized by the introduction of gap junction coupling and synchronizing dynamics, or Kv-channel properties, affecting the burst behavior of $\beta$-cells and quantitatively the intercellular calcium events. Additionally, these data demonstrated that the form of membrane potential oscillations, correlated with $\textrm{Ca}^{2+}$ concentration oscillations, is a necessary factor in $\beta$-cell calcium elevation, in addition to inter-$\beta$-cellular communications and islet synchrony. As changes in the gating of $\textrm{K}^{+}$-channels can yield an excess of large events in the patterns and activity of $[\textrm{Ca}^{2+}]_{\textrm{c}}$, and the pulse mass of insulin secretion, it will be necessary to uncover the underlying mechanisms of normal Kv-channel function for potential diabetes therapies, however, the exact reason for this remains still unclear.

\section*{Conclusion}\hypertarget{Concl}{}

Our knowledge about the differential contribution of human $\beta$-cell mass and function in hyperglycemia development and  T2D pathogenesis can provide key information for regenerating $\beta$-cell mass or preserving $\beta$-cell function. This study demonstrates that $\beta$-cell mass reduction is an important factor in $\beta$-cell dysfunction, impairment in intra-islet synchrony, and changes in the shape of electrical bursting, which cause changes in insulin secretion dynamics and insulin levels. The role of gap junction-mediated electrical coupling in affecting the behavior of intracellular $\textrm{Ca}^{2+}$ dynamics is more significant compared with both metabolic coupling and synchronous oscillations of islet activity. Our results reveal that in human $\beta$-cells lacking gap junctions modifying electrical patterns to enhance $[\textrm{Ca}^{2+}]_{\textrm{c}}$ levels and the amount of secreted insulin can arise from changes in the expression of Kv-channels, pointing towards a prominent role of Kv-channels in T2D development and therapy.

\section*{Acknowledgments}\hypertarget{Acknow}{}

The authors would like to thank Professor Azadeh Ebrahim-Habibi (Biosensor Research Center, Endocrinology and Metabolism Molecular-Cellular Sciences Institute, Tehran University of Medical Sciences) for helpful suggestions and discussions in performing and developing this work, and Professor Alessandro Loppini (Campus Bio-Medico University of Rome) for his comments on this study. The authors would also like to thank Poorya Khanizadeh for helping them to edit this manuscript. The second author is indebted to the Research Core: ‘Bio-Mathematics with computational approach’ of Tarbiat Modares University, with grant number ‘IG-39706’.

\end{document}